\documentclass{bmvc2k}

\usepackage{enumitem}

\usepackage{algpseudocode}  
\usepackage{amsmath}
\usepackage{verbatim}
\usepackage{amssymb,amsthm,array,mathtools}
\usepackage{calc}
\usepackage{tikz}
\usepackage{wrapfig}
\usepackage{threeparttable}
\usepackage{colortbl}
\usepackage{makecell}
\usepackage{diagbox}
\usepackage{graphicx}
\usepackage{subcaption}
\usepackage{epsfig}
\usepackage{times}
\usepackage{multicol}
\usepackage{multirow}
\usepackage[ruled,linesnumbered]{algorithm2e}
\usepackage{bm}
\usepackage{caption}
\usepackage{xcolor}
\usepackage{arydshln}   %\hdashline

\usepackage{nicefrac}
\usepackage{wrapfig}

\usepackage[misc]{ifsym}
\providecommand*\EnableMain[1]{}
\let\EnableMainstart=\iftrue       %正文保留
\let\EnableMainend=\fi

\providecommand*\EnableAppx[1]{}
\let\EnableAppxstart=\iftrue       %附录保留
\let\EnableAppxend=\fi

\EnableMainstart   %正文开关起作用的地方 （开始）

\title{\fontsize{16}{60}\selectfont $\;\;\;\;\;\;\;$Pay Self-Attention to Audio-Visual Navigation}

\addauthor{$\;\;\;\;\;\;\;\;\;\;\;$Yinfeng Yu}{yyf17@mails.tsinghua.edu.cn}{1, 3, $*$}
\addauthor{$\;\;\;\;\;\;\;\;\;\;\;$Lele Cao}{lele.cao@eqtpartners.com}{1, 2, $*$}
\addauthor{$\;\;\;\;\;\;\;\;\;\;\;$Fuchun Sun}{fcsun@mail.tsinghua.edu.cn}{1, \textrm{\Letter}}
\addauthor{$\;\;\;\;\;\;\;\;\;\;\;$Xiaohong Liu}{liu-xh17@mails.tsinghua.edu.cn}{1}
\addauthor{$\;\;\;\;\;\;\;\;\;\;\;$Liejun Wang}{wljxju@xju.edu.cn}{3}

\addinstitution{
    Department of Computer Science \\
    and Technology, State Key Lab on \\
    Intelligent Technology and Systems, \\
    Tsinghua University, Beijing, China
}
\addinstitution{
    Motherbrain, EQT, Stockholm, Sweden
}
\addinstitution{
    College of Information Science \\
    and Engineering, Xinjiang University, \\
    Urumqi, China
}

\runninghead{Yu, Cao, Sun, Liu, Wang}{Pay Self-Attention to Audio-Visual Navigation}

\begin{document}

\maketitle

\vspace{-12pt}
\begin{abstract}
Audio-visual embodied navigation, as a hot research topic, aims training a robot to reach an audio target using egocentric visual (from the sensors mounted on the robot) and audio (emitted from the target) input. 
The audio-visual information fusion strategy is naturally important to the navigation performance, but the state-of-the-art methods still simply concatenate the visual and audio features, potentially ignoring the direct impact of context.
Moreover, the existing approaches requires either phase-wise training or additional aid (e.g. topology graph and sound semantics). 
Up till this date, the work that deals with the more challenging setup with moving target(s) is still rare.
As a result, we propose an end-to-end framework FSAAVN (feature self-attention audio-visual navigation) to learn chasing after a moving audio target using a context-aware audio-visual fusion strategy implemented as a self-attention module.
Our thorough experiments validate the superior performance (both quantitatively and qualitatively) of FSAAVN in comparison with the state-of-the-arts, and also provide unique insights about the choice of visual modalities, visual/audio encoder backbones and fusion patterns.
\end{abstract}

%-------------------------------------------------------------------------
%===============================================================================
\begin{figure*}[ht!]
\centering
\includegraphics[width=0.72\textwidth]{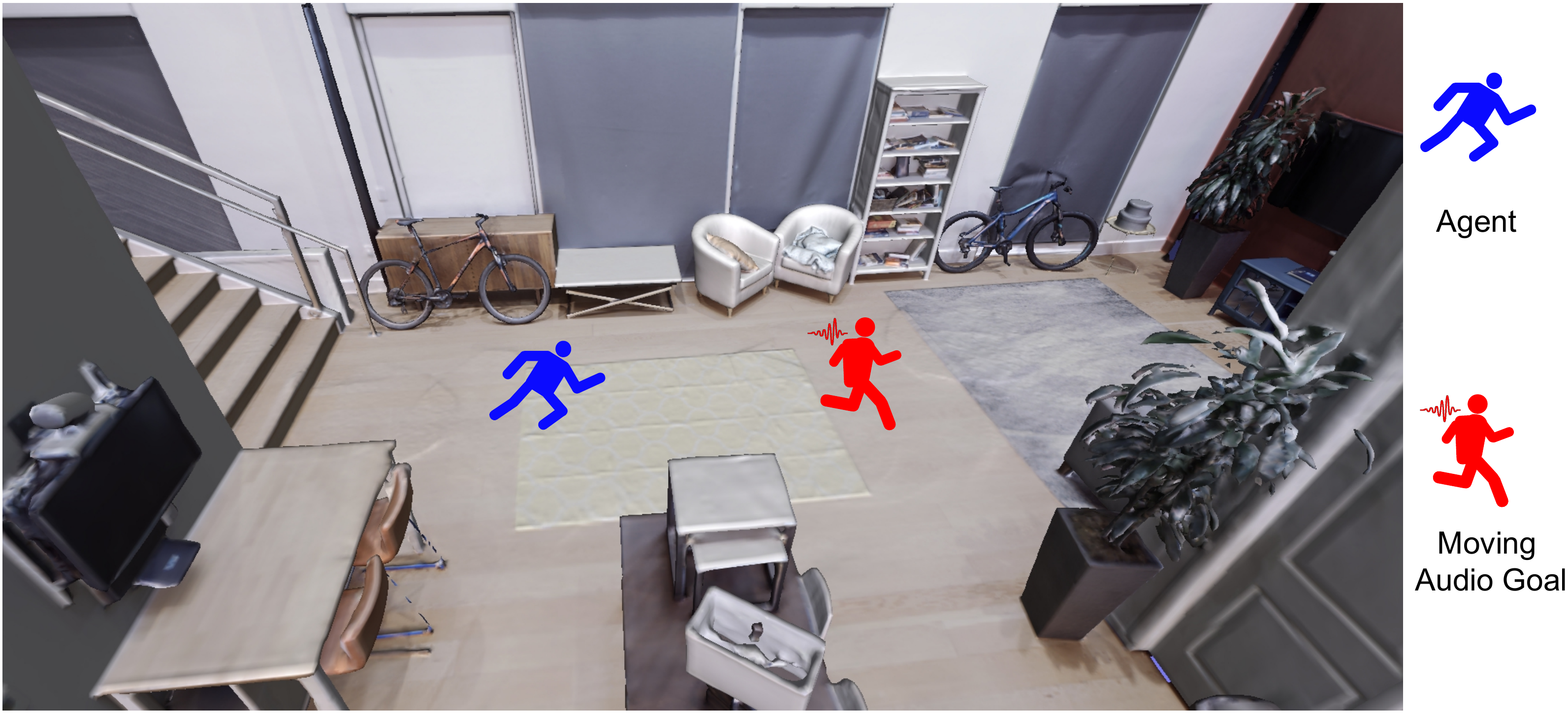}
\vspace{-3pt}
\caption{\small Audio-visual embodied navigation with a moving sound source as the target: a blue robot chases a moving target (red) that is a low-speed robot emitting sound.
}
\label{fig: title-fig}
%\vspace{-7pt}
\end{figure*}

\section{Introduction}
Embodied navigation~\cite{VisualNavigation01,VisualNavigation02, dd-ppo,cl} involves robotic agents (with egocentric observation)~\cite{visualExplor1, visualExplor2, visualExplor3, visualExplor4, navInComEnv} exploring the unknown environment~\cite{visualInteraction02, visualInteraction03, visualInteraction04} to reach target (sometimes moving) locations.
Embodied navigation has been a hotspot research topic in the broader domain of embodied intelligence~\cite{cangelosi2015embodied}.
% Agents interact with the environment in an unknown environment~\cite{visualInteraction02, visualInteraction03, visualInteraction04}, exploring through first-perspective observation~\cite{visualExplor1, visualExplor2, visualExplor3, visualExplor4, navInComEnv}, always research hotspots of embodied intelligence~\cite{VisualNavigation01,VisualNavigation02, dd-ppo,cl}.
Up till this date, most embodied navigation work relies on sensors such as vision and lidar~\cite{nnSLAM, diffSLAM}, ignoring other vital senses like hearing heavily utilized by some animals~\cite{ChristensenHY20BatVision,tracy2021catchatter, cloudNavi}.
Hearing is a unique and important sense because it is known to be both temporal and spatial informative~\cite{spatiotemporal, VisualEchoes, ImageToDepth}, enabling the visually impaired subjects to navigate properly~\cite{Wayfinding}.
Inspired by the simultaneous use of visual and hearing by animals and humans~\cite{childLearn01,childLearn02}, audio-visual assistance is believed to be beneficial to the efficiency and robustness \cite{tian2021can, SAAVN} of many different robotic tasks, such as audio-visual association~\cite{SHE}, moving vehicle tracking~\cite{gan2019self}, 
%-----------------------------------------------------------------------
% \begin{wrapfigure}{rt}{0.6\textwidth}
% \vspace{-10pt}
%   \begin{center}
%     \includegraphics[width=0.58\textwidth]{fig/fsa-title-fig-v2-cropped.pdf}
%   \end{center}
%   \vspace{-8pt}
%   \caption{\small Audio-visual embodied navigation with a moving sound source as the target: a blue robot chases a moving target (red) that is a low-speed robot emitting sound.}
%   \vspace{-7pt}
% \label{fig: title-fig}
% \end{wrapfigure}
%-----------------------------------------------------------------------
visual sound separation~\cite{gan2020music,Move2Hear}, object detection~\cite{mo-tracking}, audio-visual dereverberation \cite{av-dereverb}, audio-visual matching \cite{acoustic-matching,f-av-acousitc,chen22soundspaces2}, audio-visual floor plan reconstruction \cite{BoniardiVMCB19, purushwalkam2021audio}, and finally the main focus of this work: audio-visual embodied navigation~\cite{LLA,AV-WaN,SAVi,SoundSpaces,SAAVN,CMHM}.

Among the state-of-the-art audio-visual navigation researches, some \cite{LLA} require multiple sequential steps to reach the goal; some \cite{LLA,AV-WaN} need to build topology graph; and some \cite{SAVi,fallen-obj-avn} rely on sound meta information. 
Unfortunately, all of these methods can only deal with single sound target that stays at the same position throughout the entire navigation task. 
Most recently, the authors of \cite{SAAVN} and \cite{CMHM} experiment with the multiple sound sources and moving sound source, respectively.
However, they simply apply concatenation to fuse the visual and audio information, as shown in Fig.\ref{fig: compare-ideal}(a).
Under this status-quo, we propose an end-to-end framework -- Feature Self-Attention Audio-Visual Navigation (FSAAVN), which supports chasing a moving sound target without the need of topology graph or sound meta information.
More importantly, as shown in Fig.\ref{fig: compare-ideal}(d), we propose a novel audio-visual fusion module FSA (feature self-attention) to learn a context-aware strategy to determine the relative contribution of each modal in real-time.
The main contributions of this work are:
\vspace{-3pt}
\begin{itemize}
\setlength\itemsep{-2pt}
\item we propose a end-to-end framework (FSAAVN) to address a currently under-researched problem: audio-visual navigation to chase a moving sound target;
\item we design a novel audio-visual fusion module (FSA) to learn a context-aware strategy to determine the relative contribution of each modal in real-time;
\item we experimentally benchmark our approach towards the state-of-the-arts in 3D environments, showing the superior performance of FSAAVN;
\item the thorough comparison of different variants of the fusion module (Fig.\ref{fig: compare-ideal}) and visual/audio encoder\footnote{The considered encoders: CNN (convolution neural network), ViT (vision transformer)~\cite{ViT}, Capsule~\cite{Capsules}. } provides useful insights for future practitioners in this field.
	
\begin{figure*}[ht!]
\centering
\includegraphics[width=0.95\textwidth]{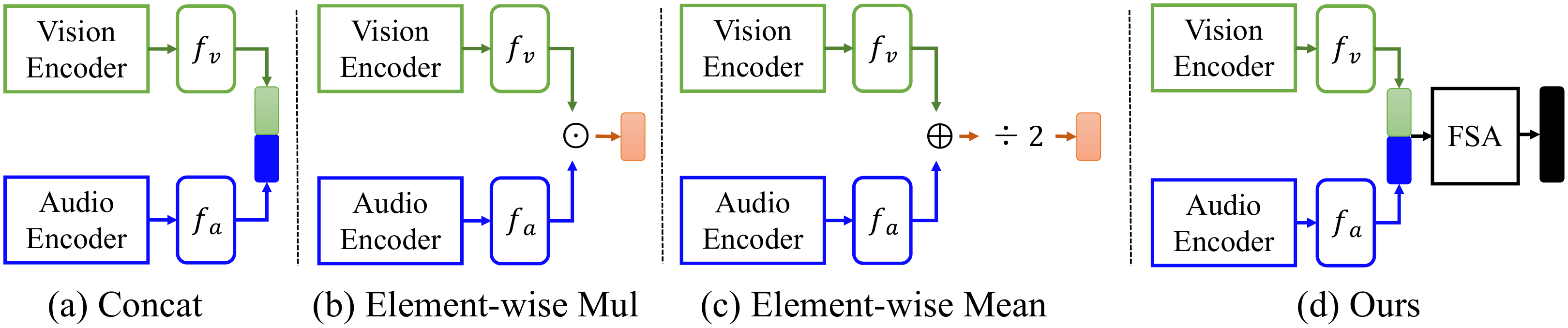}
\vspace{-3pt}
\caption{\small High-level illustration of different feature fusion methods in audio-visual navigation.
}
\label{fig: compare-ideal}
\vspace{-7pt}
\end{figure*}
	
\end{itemize}

\section{Related Work } \label{sec:relatedwork}

\begin{figure*}[ht]
\centering
\includegraphics[width=\textwidth]{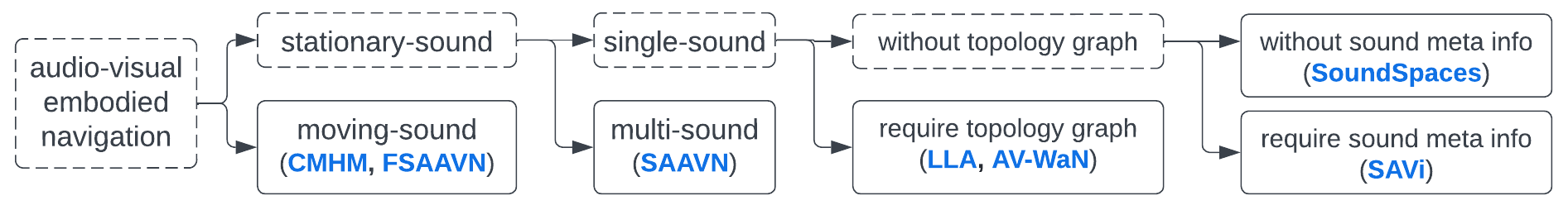}
\vspace{-15pt}
\caption{\small The landscape of the related work addressing audio-visual embodied navigation.
}
\label{fig: survey-landscape}
\vspace{-3pt}
\end{figure*}

Audio-visual embodied navigation is largely grouped into two categories in accordance with the behavior of sound source(s): static-sound and moving-sound sources.
We visually illustrate the landscape of our literature survey in Fig.\ref{fig: survey-landscape}.
\vspace{-6pt}

{\bf Stationary-sound} refers to an environment where the location of the sound source is preset and remains the same. 
The number of sound sources could be one ({\it single-sound})~\cite{LLA,AV-WaN,SAVi,SoundSpaces} or many ({\it multi-sound})~\cite{SAAVN}. 
To this date, the majority researches of audio-visual navigation have been carried out in {\it single-sound} environment, such as LLA (look, listen and act)~\cite{LLA}, SoundSpaces~\cite{SoundSpaces}, AV-WaN (audio-visual waypoint navigation)~\cite{AV-WaN}, SAVi (semantic audio-visual navigation)~\cite{SAVi}.
LLA~\cite{LLA} is a classical phase-wise navigation solution that needs to build a topological graph to aid shortest path planning.
SoundSpaces~\cite{SoundSpaces} is the first end-to-end approach that does not rely on any topology graph or sound meta information (e.g. the category of the sound source: telephone, doorbell, alarm, etc).
AV-WaN~\cite{AV-WaN} predicts waypoints (represented as a topological graph) to improve long-distance navigation performances.
The authors of SAVi~\cite{SAVi} address the situation where the sound is not emitted continuously by incorporating the sound meta information. 
In the {\it multi-sound} case, SAAVN (sound adversarial audio-visual navigation)~\cite{SAAVN} propose and end-to-end framework to deal with acoustically complex environments where the target sound (usually only one source) is mixed up with other noisy sounds (usually more than one source).

{\bf Moving-sound}, as the name implies, refers to a sound source that keeps changing its position, and the navigation target is that moving sound source. 
To the best of our knowledge, CMHM (catch me if you hear me)~\cite{CMHM} is the first and only published work to tackle the moving-sound problem within unexplored environments.
This work (FSAAVN) is closely related to CMHM with significant advancements mainly in audio-visual fusion strategy.

{\bf Audio-visual fusion}: normally, the audio and visual inputs are encoded (using different encoders) into audio and visual feature vectors, respectively.
They need to be fused before fed to the downstream neural networks.
Unfortunately, all of the aforementioned approaches simply concatenate them to form a fused feature vector, as shown in Fig.\ref{fig: compare-ideal}(a). 
In this work, we attempt to solve the moving-sound navigation problem using a more advanced audio-visual fusion mechanism: Feature Self-Attention (FSA) as illustrated in Fig.\ref{fig: compare-ideal}(d).

\vspace{-4pt}
\section{Setting the Stage and Goal}
\label{section:stage_goal}
\vspace{-2pt}
In 3D environments demonstrated in Fig.\ref{fig: title-fig}, a robot learns to chase after and catch up with a moving sound target. 
To concretize the research, we adopt the commonly used 3D environments collected using the SoundSpaces platform~\cite{SoundSpaces} and Habitat simulator~\cite{habitat-sim}.
They are publicly available as several datasets: Replica~\cite{replica}, Matterport3D~\cite{matterport3d} and SoundSpaces (audio) ~\cite{SoundSpaces}.
Replica contains 18 environments in the form of grids (with a resolution of 0.5 meter) constructed from accurate scans of apartments, offices and hotels. 
Matterport3D has 85 scanned grids (1 meter resolution) of indoor environments like personal homes.
In SoundSpaces, a sound source emits omnidirectional sound that is convolved with the corresponding binaural RIR (room impulse response);
the convolved result is a binaural environmental response that is received by the navigating robot from its facing direction.

\begin{figure*}[t!]
\centering
\includegraphics[width=0.92\textwidth]{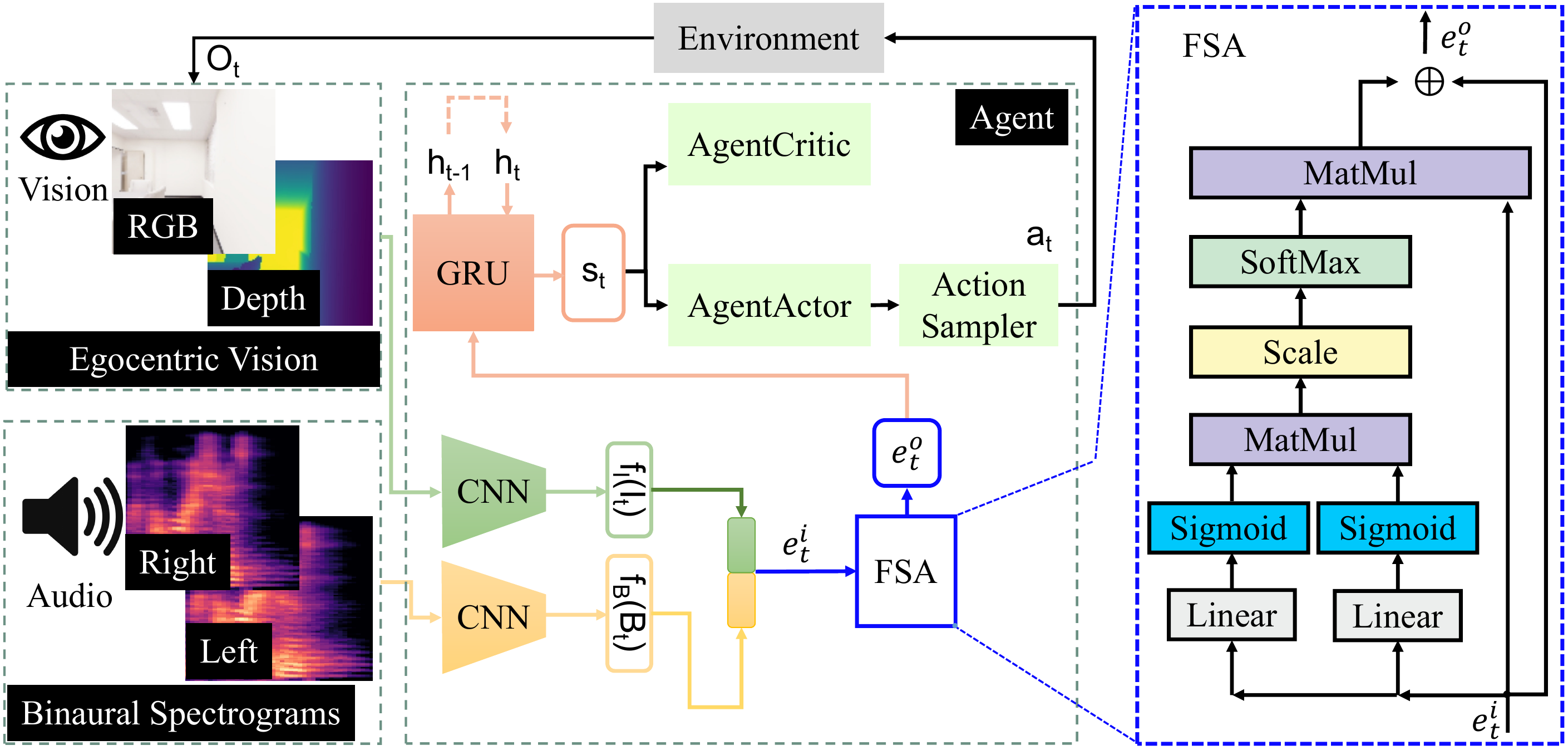}
\caption{\small The overview of FSAAVN: Feature Self-Attention Audio-Visual embodied Navigation. 
%The agent first encode observations to obtain the input embedding vector $e^{i}_{t}$, and transform the input embedding vector $e^{i}_{t}$ to the output embedding vector $e^{o}_{t}$ by Feature Self-Attention,  and then learn state vector $s_{t}$. Then, state vector $s_{t}$ are fed to actor-critic networks, which predict the next action $a_{t}$. The agent receive its reward from the environment.
}
\label{fig: network}
\vspace{-9pt}
\end{figure*}

At each step $t$ (cf. Fig.\ref{fig: network}), the robot receives the current observation $O_{t}=(I_{t}, B_{t})$, 
where $I_{t}$ denotes the current visual input that can be RGB (128$\times$128$\times$3 pixels) and/or depth (with a dimension of 128$\times$128$\times$1) image\footnote{Both RGB and depth images capture the 90-degree field of view in front of the navigating robot.}, 
$B_{t}$ represents the received binaural target sound spectrum\footnote{\scriptsize The audio spectrum is prepared in the same way as \cite{SoundSpaces, AV-WaN}: we compute the short-time Fourier transform with a window length of 512 samples and a hop length of 160 samples, corresponding to a duration of 12ms and 32ms in Replica \cite{replica} and Matterport3D \cite{matterport3d}.
Taking an aggregate of 1 sec gives matrices with dimensions of 257$\times$257 (Replica) and 257$\times$101 (Matterport3D). 
We further down-sample each matrix by a factor of 4, and stack the ones from left and right ``ear'' in a third channel, resulting in the final spectrum matrices in dimensions of 65$\times$69$\times$2 (Replica) and 65$\times$26$\times$2 (Matterport3D).} 
consisting of the audio signal from the left and right ``ears''.
Although there exists a {\it navigability graph} (with nodes and edges) of the environment,
this graph is hidden from the robot, hence it must learn from the accumulate observations $O_{t}$ to understand the geometry of the scene.
At each step, the agent at a certain node A can only move to another node B in the navigability graph if 1) an edge is connecting both nodes, and 2) the robot is facing node B.
The viable robotic action space is defined as $\mathcal{A}$ =\{\verb|MoveForward|, \verb|TurnLeft|, \verb|TurnRight|, \verb|Stop|\}, where the \verb|Stop| action should be executed when the robot and the moving audio target are on the same node in the navigability graph.
The overall goal of the navigating robot is to {\bf learn a policy to catch up with the moving audio target as fast as possible using both the visual and audio input}.

One might ask about the behavior of the moving audio target, we adopt the same approach used in CMHM~\cite{CMHM}.
In the beginning, the sound source randomly selects a destination position to move to; 
the selected destination position must have a traversable path to the current position of the sound source.
Then, the sound source starts to follow the shortest path to reach (step by step) the designation position.
However, at each step, the robot may only turn left/right while the sound source always match to the next node in the navigability graph.
As a result, each move step of the sound source occurs with only 30\% probability, so that the robot has a possibility to eventually catch up with the moving sound source.
Once the sound source reaches the current destination position, a new destination position is randomly selected again so that the sound source will continue moving (in exactly the same manner described previously) to the new destination until the robot catches it.

%\vspace{-4pt}
\section{The Proposed Approach} 
\label{sec:approach}
%\vspace{-2pt}
We formalize the problem as a reinforcement learning task in which the navigating robot learns a policy to catch up with a moving audio target quickly in an unknown environment.
Our solution to the problem is abbreviated FSAAVN (feature self-attention audio-visual navigation). 
FSAAVN is composed of four main parts (Fig.\ref{fig: network})
Specifically, given the egocentric vision and audio input, 
our model 1) encodes visual/audio inputs into visual/audio features using {\bf CNNs} (convolution neural networks); 
2) fuse the visual and audio features with {\bf FSA} (feature self-attention) producing the fused audio-visual embedding, 
3) transform a series of such embeddings into a temporal-ware state representation using a {\bf GRU} (gated recurrent unit);
and finally 4) use an {\bf actor-critic} network to perform action prediction, evaluation, and optimization.
The robot agent repeat this process until it catches the moving audio target.
We will introduce each part consecutively in the upcoming paragraphs.

As introduced in Section~\ref{section:stage_goal}, the robot receives an audio-visual observation $O_{t}=(I_{t}, B_{t})$ at the $t$-th step.
The visual ($I_{t}$) part is encoded into visual feature vector using a CNN encoder: $f_{I}(I_{t})$.
The audio feature vector is obtained in the same way using a different CNN encoder: $f_{B}(B_{t})$.
Visual and auditory CNN encoders are constructed in the same way (from the input to output layer): \verb|Conv8x8|, \verb|Conv4x4|, \verb|Conv3x3| and a 512-dim linear layer; 
ReLU activations are added between any two neighboring layers.
We denote the concatenation of visual and audio features as $e^{i}_{t}=[ f_{I}(I_{t}),\,f_{B}(B_{t})]$. 

To determine the relative contribution of each modal in real-time according to the varying context, we design a trainable audio-visual fusion mechanism (i.e. FSA) to transform the encoded features $e^{i}_{t}$ to a fused embedding vector $e^{o}_{t}$:
\begin{equation}
	e^{o}_{t} = \operatorname{softmax}\left(\frac{\operatorname{sigmoid}(W_{1}Q) \operatorname{sigmoid}(W_{2}K)^{T}}{\sqrt{d}}\right) V \oplus V,
	\quad \text{s.t.} \;
	Q=K=V=e^{i}_{t},
\end{equation}
%-------------------------
where $Q,\,K,\,V$ are the query, key, and value input to the FSA module; $W_{1}$ and $W_{2}$ are both weight matrices to be optimized; $d$ is a scalar factor with a value of 256; and $e^{o}_{t}$ is the resulting fused embedding vector. We collectively denote the encoder and FSA weights as $\mathbf{W}$ hereafter for simplicity. The right part of Fig.\ref{fig: network} can be referred to for more details.

A bidirectional GRU (with one 512-dim hidden layer) is applied to further transform a series of fused embeddings (i.e. $e^{o}_{1}\ldots e^{o}_{t}$) into a temporal-aware state representation $s_{t}$. Concretely, at time $t$, the GRU cell takes in both the current embedding $e^{o}_{t}$ and the previous cell state $h_{t-1}$ to produce $s_{t}$ and $h_{t}$. Essentially, $s_{t}= GRU(e^{o}_{t},\,h_{t-1})$.

The state vectors (i.e. $s_{1}\ldots s_{t}$) is then fed to an actor-critic network to 1) predict the conditioned action probability distribution $\pi_{\theta_1}(a_t|s_{t})$, 
and 2) estimate the state value $V_{\theta_2}(s_{t})$.
The actor and critic are implemented with a single linear layer parameterized by $\theta_1$ and $\theta_2$, respectively.
For the sake of conciseness, we use $\boldsymbol\theta$ to denote the compound of $\theta_1$ and $\theta_2$ hereafter.
The action sampler in Fig.\ref{fig: network} samples the actual action (i.e. $a_t$) to execute from $\pi_{\theta_1}(a_t|s_{t})$. 
The training aims to maximise the expected discounted return $\Re$:
\begin{equation} 
\label{main: exp-return}
    \Re = \mathbb{E}_{\pi}\left[\textstyle\sum_{t=1}^{T} \gamma^{t} r\left(s_{t-1}, a_{t}\right)\right],
\end{equation}
where $\gamma$ is a discount factor; $T$ is the maximum number of time steps; and $\pi$ is the policy of the robot agent.
$r\left(s_{t-1}, a_{t}\right)$ is the reward given by the environment at the time step $t$.
The reward is calculated based on three simple rules: (1) \verb|+10| point when the robot successfully reaches the target and executes the \verb|Stop| action, (2) \verb|+0.25| point when the Manhattan distance between the robot and target is reduced, and (3) a time penalty of \verb|-0.01| on each action performed to encourage navigation efficiency.
Proximal Policy Optimization (PPO)~\cite{ppo} is adopted in this work to optimize \eqref{main: exp-return}.
The entire procedure is described in Algorithm~\ref{algorithm:fsaavn}.
%as pseudo code.

\begin{algorithm}[ht!]
	\caption{FSAAVN: feature self-attention audio-visual navigation}
	\label{algorithm:fsaavn}
	\KwData{
		Environment $\mathcal{E}$,
		stochastic policies $\pi$,
		initial actor-critic weights $\boldsymbol\theta_0$,
		initial encoder and FSA weights $\mathbf{W}_0$,
		\# updates $M$, 
		\# episode $N$,
		max time steps $T$.
	}
	\KwResult{
		Trained weights: $\boldsymbol\theta_{M}$ and $\mathbf{W}_M$
	}
	\For {$i$=1, 2, ... $M$}
	{
		// Run policy $\pi_{\boldsymbol\theta_{i-1}}$ in environment for $N$ episodes $T$ time steps   \;
		$\{(o_{t}, \,h_{t-1}, \,a_{t}, \,r_{t})_{i}\}_{t=1}^{T} \leftarrow \text{roll}(\mathcal{E}, \pi_{\boldsymbol\theta_{i-1}}, T)$  at $i$-th update \;
		% Compute advantage estimates $\hat{A}_{1},\cdots,\hat{A}_{T}$ \;
		Compute advantage estimates \;
		// Optimize w.r.t. $\boldsymbol\theta$ and $\mathbf{W}$ \;
		$\boldsymbol\theta_{i}, \mathbf{W}_i \leftarrow$ new $\boldsymbol\theta$ and $\mathbf{W}$ from PPO algorithm w.r.t. maximizing Equation~\eqref{main: exp-return}\;  
	}
\end{algorithm}

% \subsection{Why not replace GRU with Attention}
While the self-attention mechanism (i.e. FSA) over the latent dimension is effective (experimentally validated in Section~\ref{sec: experiments}), why don't we choose to adopt attention over the temporal dimension (i.e. replace the GRU with Attention)?
The reason is two fold: 
\begin{itemize}
    \item The input data for audio-visual navigation is different from the typical input we see in NLP (natural language processing) and CV (computer vision) tasks. At any time point, the observation is always incomplete, meaning the future states are unknown. As a result, learning the temporal attention can be extremely unstable.
    \item The maximum number of navigation steps is 500 (or one can choose a much larger number), which leads to a spacial-temporal transformer model with high capacity and expressivity, hence requiring large amount of samples/episodes to converge.
\end{itemize}
%\clearpage
%\vspace{-4pt}
\section{Experiments} \label{sec: experiments}
%\vspace{-2pt}
We carry out experiments on the environments and datasets described in section~\ref{section:stage_goal}.
FSAAVN is benchmarked towards several state-of-the-art baselines: {\bf SoundSpaces}~\cite{SoundSpaces}, {\bf SoundSpaces-EMul}, {\bf SoundSpaces-EM}, {\bf CMHM}~\cite{CMHM}, and {\bf AV-WaN}~\cite{AV-WaN}.
SoundSpaces-EMul/-EM is the extension of basic SoundSpaces (concatenation fusion) with different audio-visual feature fusion methods: ``EMul'' stands for element-wise multiplication and ``EM'' is element-wise mean, as shown in Fig.\ref{fig: compare-ideal}(b,c).
Based on the widely used SPL (success path length) metric~\cite{metric-SPL}, we calculate several evaluation metrics, such as {\bf SPLT} (SPL for tracking), {\bf SSPLT} (soft SPLT), and {\bf SRT} (success rate for tracking).
%\vspace{-3pt}
\begin{equation}
\mathrm{SPLT}\!=\!\frac{1}{N}\sum_{i=1}^{N}\frac{S_i\cdot l_i}{\mathrm{max}(p_i,l_i)},\quad
\mathrm{SSPLT}\!=\!\frac{1}{N}\sum_{i=1}^{N} \frac{l_i\cdot\max(0,1\!-\!\frac{d^{a}_{i}}{d_{i}})}{\mathrm{max}(p_i,l_i)},\quad
\mathrm{SRT}\!=\!\frac{1}{N}\sum^{N}_{i=1}S_{i},
\label{eq:metrics}
%\vspace{-5pt}
\end{equation}
where $N$ is the number of episodes; $S_{i}$ is a binary indicator of success in the $i$-th episode; $p_i$ stands for the length of the executed path; $l_i$ is the length of the shortest path from the robot's start position to the target's final position. When computing SSPLT by the end of the $i$-th episode, $d^{a}_{i}$ is the robot's distance to the target, $d_{i}$ is the distance from the robot's start position to the target's final position.
\EnableMainstart 
    \EnableAppxstart
        %正文、附录的引用正常
        For the definition of other evaluation metrics used in this research, please refer to Appx.\ref{appx: metrics}.
    \else
        %正文，硬写对附录的引用
        For the definition of other evaluation metrics used in this research, please refer to Appx.B.1~\cite{fsaavn}.
    \EnableAppxend
\else
    %附录，硬写对正文的引用
    % appx hard cite main
\EnableMainend
For all metrics, completion of an episode indicates that the robot either catches the target in less than 500 steps, or selects the stop action precisely at the location of the moving target. 
The reported metric values are averaged over 5 trials.

We train our model with Adam (to optimize an entropy loss on the policy distribution) with a learning rate of $2.5 \times 10^{-4}$ with a limit of time horizon corresponding to 500 actions in a scene.
We train the framework for $40M$ steps on Replica and $60M$ on Matterport3D, which amounts to 200 and 320 GPU hours, respectively.
Since there are 102 sounds from SoundSpaces, we test three different sound source splittings.
(1) {\bf Telephone}: the target sound source (telephone) is the same in the training, validation and testing sets;
(2) {\bf Multiple heard}: all 102 sounds exist in three sets; 
(3) {\bf Multiple unheard}: the 102 sounds are divided into non-overlapping 73/11/18 splits for train/validation/test.

%\clearpage
\subsection{Overall performance comparison}
\EnableMainstart 
    \EnableAppxstart
        %正文、附录的引用正常
        We generally discover that depth works the best with sound input (cf. Table~\ref{tab: fusion-visual} and Appx.\ref{appx: compare-baselines}) that coincide the results in \cite{SoundSpaces}.
    \else
        %正文，硬写对附录的引用
        We generally discover that depth works the best with sound input (cf. Table~\ref{tab: fusion-visual} and Appx.B.3 ~\cite{fsaavn}) that coincide the results in \cite{SoundSpaces}.
    \EnableAppxend
\else
    %附录，硬写对正文的引用
    % appx hard cite main
\EnableMainend
Therefore, in Table~\ref{tab: fusion}, we only illustrate the metrics obtained using depth and sound input.
It shows that FSAAVN using FSA fusion constantly performs the best (in boldface) on both datasets in all splitting settings.

\begin{table*}[t!]
\addtolength{\tabcolsep}{-5pt}
\renewcommand{\arraystretch}{1.2}
\centering
\caption{\small
	Overall performance comparison (STDEV$\leq$0.01) using depth and sound input.
}
\label{tab: fusion}
\vspace{0.1in}
\resizebox{1\linewidth}{!}{
\begin{tabular}{l|l|ccc|ccc|ccc|ccc|ccc|ccc} 
\hline
\multirow{4}{*}{Model} & \multirow{4}{*}{Fusion} & \multicolumn{9}{c|}{Replica}  & \multicolumn{9}{c}{Matterport3D}      \\ 
\cline{3-20}
             &      & \multicolumn{3}{c|}{Telephone} & \multicolumn{3}{c|}{Multiple heard} & \multicolumn{3}{c|}{Multiple unheard} 
                    & \multicolumn{3}{c|}{Telephone} & \multicolumn{3}{c|}{Multiple heard} & \multicolumn{3}{c}{Multiple unheard}  \\ 
\cline{3-20}
             &         & SPLT               & SSPLT             & SRT                
                       & SPLT               & SSPLT             & SRT                          
                       & SPLT               & SSPLT             & SRT                   
                       & SPLT               & SSPLT             & SRT                     
                       & SPLT               & SSPLT             & SRT                          
                       & SPLT               & SSPLT             & SRT                   \\ 
% \cline{3-20}
             &         & ($\uparrow$)       & ($\uparrow$)      & ($\uparrow$)                 
                       & ($\uparrow$)       & ($\uparrow$)      & ($\uparrow$)                     
                       & ($\uparrow$)       & ($\uparrow$)      & ($\uparrow$)               
                       & ($\uparrow$)       & ($\uparrow$)      & ($\uparrow$)                  
                       & ($\uparrow$)       & ($\uparrow$)      & ($\uparrow$)                       
                       & ($\uparrow$)       & ($\uparrow$)      & ($\uparrow$)                \\ 
\hline
FSAAVN   &  FSA      & \textbf{0.541} & \textbf{0.635} & \textbf{0.925} & \textbf{0.438} & \textbf{0.541} & \textbf{0.812} & \textbf{0.182} & \textbf{0.316} & \textbf{0.358} & \textbf{0.520} & \textbf{0.585} & \textbf{0.832} & \textbf{0.438} & \textbf{0.496} & \textbf{0.844} & \textbf{0.207} & \textbf{0.299} & \textbf{0.391}    \\
 %--------------------------------------------------------------------------------------------------------------------
SoundSpaces        &  Concat   & 0.531 & 0.604 & 0.892 & 0.354 & 0.462 & 0.764 & 0.152 & 0.255 & 0.317 & 0.454 & 0.511 & 0.797 & 0.431 & 0.475 & 0.818 & 0.180 & 0.254 & 0.350    \\
 %--------------------------------------------------------------------------------------------------------------------                       
SoundSpaces-      &  EMul   & 0.493 & 0.597 & 0.861 & 0.430 & 0.522 & 0.770 & 0.168 & 0.304 & 0.326 & 0.457 & 0.523 & 0.801 & 0.433 & 0.481 & 0.821 & 0.182 & 0.258 & 0.355    \\
 %--------------------------------------------------------------------------------------------------------------------                       
SoundSpaces-       &  EM  & 0.487 & 0.592 & 0.816 & 0.435 & 0.531 & 0.796 & 0.154 & 0.258 & 0.319 & 0.481 & 0.543 & 0.817 & 0.435 & 0.492 & 0.832 & 0.183 & 0.266 & 0.375    \\
 %--------------------------------------------------------------------------------------------------------------------                       
CMHM       &  Concat   & 0.335 & 0.338 & 0.791 & 0.259 & 0.302 & 0.692 & 0.121 & 0.202 & 0.314 & 0.114 & 0.125 & 0.606 & 0.086 & 0.099 & 0.528 & 0.052 & 0.085 & 0.267    \\
 %--------------------------------------------------------------------------------------------------------------------                       
AV-WaN        &  Concat   & 0.218 & 0.224 & 0.764 & 0.220 & 0.271 & 0.533 & 0.010 & 0.189 & 0.233 & 0.111 & 0.114 & 0.409 & 0.012 & 0.034 & 0.093 & 0.010 & 0.043 & 0.057    \\
 %--------------------------------------------------------------------------------------------------------------------                       
\hline
\end{tabular}
}
\end{table*}

% %-----------------------------------------------
\begin{figure*}[t!]
	\centering
	\includegraphics[width=\textwidth]{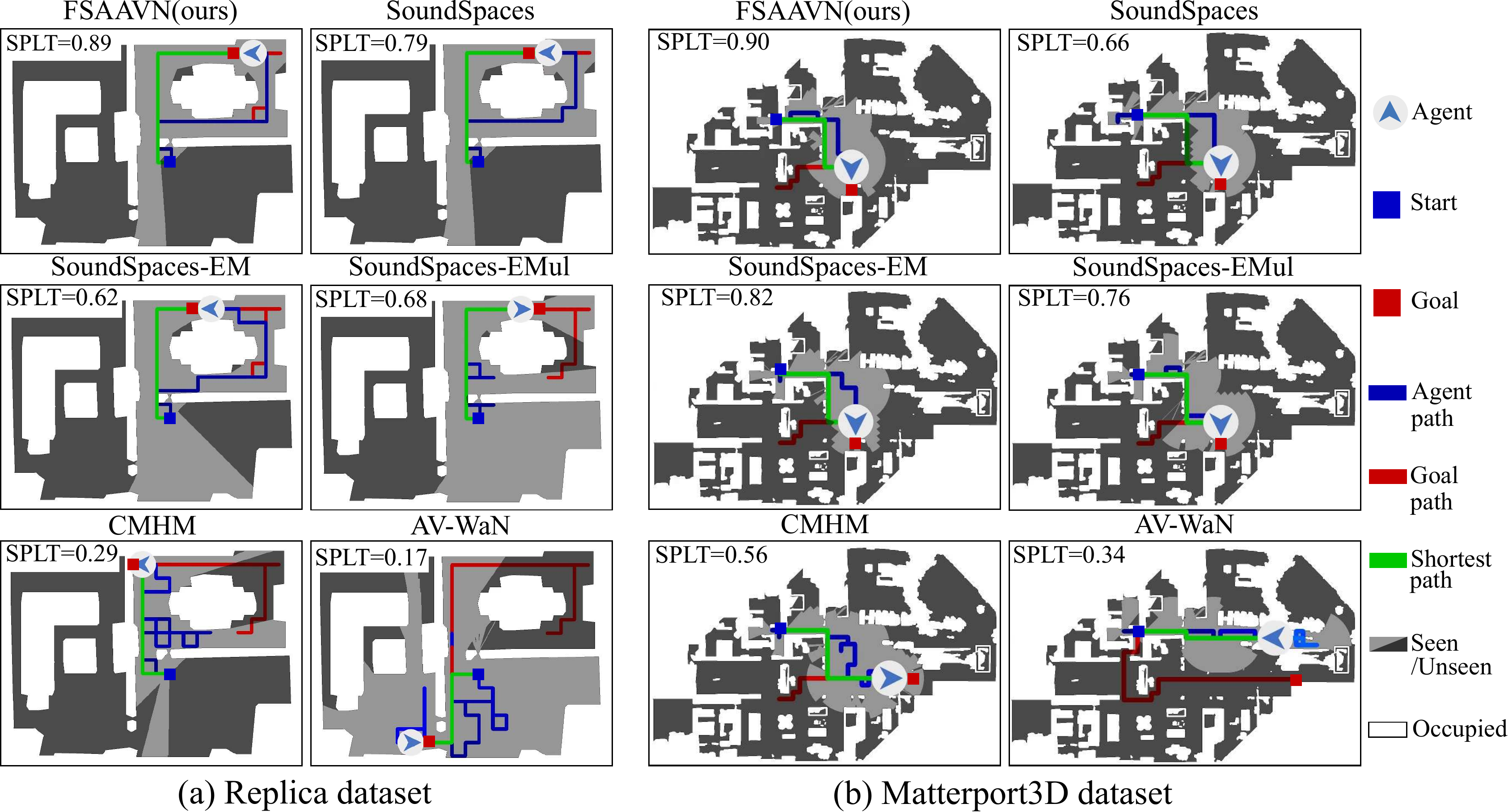}
	\vspace{-10pt}
	\caption{\small Navigation trajectories by the end of a particular episode from (a) Replica and (b) Matterport3D dataset. Higher SPLT values and shorter blue paths indicate better performances.
	}
	\label{fig: traj}
	\vspace{-3pt}
\end{figure*}
%-----------------------------------------------

To gain a qualitative and vivid impression of robot behavior from different approaches, 
we visualize the recorded navigation trajectories (using depth and sound input) on a scene map from Replica and Matterport3D datasets (Fig.\ref{fig: traj}).
On both datasets, we largely see that FSAAVN often executes a shorter trajectory and a higher SPLT value, implying a better performance than other methods.

\subsection{Comparing different visual modalities}
\label{sec: ablation-visual-modality}
There can be different types of egocentric visual input such as RGB, depth and RGBD, where RGBD is a combination of RGB and depth in different channels.
To understand the relative effectiveness of these visual modalities when fused with audio input, we compare the navigation performance between FSAAVN and SoundSpaces using RGB, depth, RGBD and blind (no visual input) in addition to the auditory input from the moving target.
From Table~\ref{tab: fusion-visual}, we can concluded that: 
1) among all tested modalities, depth alone achieves the best performance that is consistent with the conclusion in \cite{SoundSpaces} when tackling with a stationary sound source; 
2) with the same modality, FSA performs better than simple concatenation;
3) in the case of blind, FSA seems slightly worse than concatenation, probably caused by FSA's higher flexibility and complexity than concatenation.
\begin{table*}[t!]
\addtolength{\tabcolsep}{-5pt}
\renewcommand{\arraystretch}{1.2}
\centering
\caption{\small
	Performance Comparison (STDEV$\leq$0.01) of different vision modalities.
}
\label{tab: fusion-visual}
\vspace{0.1in}
\resizebox{1\linewidth}{!}{
\begin{tabular}{l|l|l|ccc|ccc|ccc|ccc|ccc|ccc} 
\hline
\multirow{4}{*}{Model} & \multirow{4}{*}{Fusion} & \multirow{4}{*}{Vision}  & \multicolumn{9}{c|}{Replica}  & \multicolumn{9}{c}{Matterport3D}      \\ 
\cline{4-21}
    &       &      & \multicolumn{3}{c|}{Telephone} & \multicolumn{3}{c|}{Multiple heard} & \multicolumn{3}{c|}{Multiple unheard} 
                    & \multicolumn{3}{c|}{Telephone} & \multicolumn{3}{c|}{Multiple heard} & \multicolumn{3}{c}{Multiple unheard}  \\ 
\cline{4-21}
    &       &          & SPLT               & SSPLT             & SRT                
                       & SPLT               & SSPLT             & SRT                          
                       & SPLT               & SSPLT             & SRT                   
                       & SPLT               & SSPLT             & SRT                     
                       & SPLT               & SSPLT             & SRT                          
                       & SPLT               & SSPLT             & SRT                   \\ 
% \cline{3-20}
    &       &          & ($\uparrow$)       & ($\uparrow$)      & ($\uparrow$)                 
                       & ($\uparrow$)       & ($\uparrow$)      & ($\uparrow$)                     
                       & ($\uparrow$)       & ($\uparrow$)      & ($\uparrow$)               
                       & ($\uparrow$)       & ($\uparrow$)      & ($\uparrow$)                  
                       & ($\uparrow$)       & ($\uparrow$)      & ($\uparrow$)                       
                       & ($\uparrow$)       & ($\uparrow$)      & ($\uparrow$)                \\ 
\hline
%--------------------------------------------------------------------------------------------------------------------
 FSAAVN   &  FSA     & Depth     & \textbf{0.541} & \textbf{0.635} & \textbf{0.925} & \textbf{0.438} & \textbf{0.541} & \textbf{0.812} & \textbf{0.182} & \textbf{0.316} & \textbf{0.358} & \textbf{0.520} & \textbf{0.585} & \textbf{0.832} & \textbf{0.438} & \textbf{0.496} & \textbf{0.844} & \textbf{0.207} & \textbf{0.299} & \textbf{0.391}    \\
 %--------------------------------------------------------------------------------------------------------------------
SoundSpaces        &  Concat   & Depth    & 0.531 & 0.604 & 0.892 & 0.354 & 0.462 & 0.764 & 0.152 & 0.255 & 0.317 & 0.454 & 0.511 & 0.797 & 0.431 & 0.475 & 0.818 & 0.180 & 0.254 & 0.350    \\
 %--------------------------------------------------------------------------------------------------------------------    
  %--------------------------------------------------------------------------------------------------------------------
\hdashline
FSAAVN &  FSA     & RGBD    &  0.532 & 0.611 & 0.837 &  0.402 & 0.485 & 0.792 &  0.185 & 0.285 & 0.349 & 0.454 & 0.510 & 0.834 & 0.440 & 0.492 & 0.827 & 0.191 & 0.281 & 0.373     \\
SoundSpaces    &  Concat  & RGBD    &  0.527 & 0.605 & 0.835 &  0.393 & 0.475 & 0.756 &  0.182 & 0.276 & 0.339 & 0.435 & 0.502 & 0.798 & 0.412 & 0.469 & 0.809 & 0.186 & 0.277 & 0.369      \\
 %-------------------------------------------------------------------------------------------------------------------- 
%--------------------------------------------------------------------------------------------------------------------
\hdashline
FSAAVN &  FSA     & RGB    & 0.530 & 0.601 & 0.872 & 0.413 & 0.500 & 0.767 & 0.166 & 0.295 & 0.305 &  0.449 & 0.505 & 0.820 & 0.393 & 0.453 & 0.781 & 0.196 & 0.270 & 0.417      \\
SoundSpaces    &  Concat  & RGB    & 0.522 & 0.593 & 0.829 & 0.386 & 0.477 & 0.741 & 0.140 & 0.260 & 0.267 & 0.397 & 0.451 & 0.815 & 0.371 & 0.429 & 0.772 & 0.193 & 0.269 & 0.375    \\
 %--------------------------------------------------------------------------------------------------------------------    
 %--------------------------------------------------------------------------------------------------------------------
\hdashline
FSAAVN &  FSA     & Blind    & 0.470 & 0.544 & 0.833 & 0.328 & 0.425 & 0.703 & 0.141 & 0.229 & 0.294 &  0.369 & 0.424 & 0.787 & 0.339 & 0.387 & 0.766 & 0.162 & 0.241 & 0.356      \\
SoundSpaces    &  Concat  & Blind    & 0.472 & 0.545 & 0.839 & 0.334 & 0.425 & 0.725 & 0.142 & 0.229 & 0.331 &  0.385 & 0.443 & 0.790 & 0.319 & 0.372 & 0.724 & 0.163 & 0.257 & 0.370     \\
 %--------------------------------------------------------------------------------------------------------------------                    
\hline
\end{tabular}
}

\end{table*}

\subsection{Comparing different visual/audio encoders}
Practically, nothing stops us from choosing a different visual/audio encoder other than CNN.
Nowadays, the common options are ViT (vision transformer)~\cite{ViT} and Capsule~\cite{Capsules}.
Here, we will compare the audio-visual navigation performance using different visual/audio encoder backbones (i.e. CNN, Capsule and ViT-based).
For ViT, we test two variants: ViT-V and ViTScratch-V, where the former is initialised with pretrained weights while the latter is trained from random weights.
\EnableMainstart 
    \EnableAppxstart
        %正文、附录的引用正常
        The implementation details of these encoders can be found in Appx.\ref{appx: encoders}.
    \else
        %正文，硬写对附录的引用
        The implementation details of these encoders can be found in Appx.A~\cite{fsaavn}.
    \EnableAppxend
\else
    %附录，硬写对正文的引用
    % appx hard cite main
\EnableMainend
Table~\ref{tab: fusion-encoder} shows the results of using different encoders while keeping the other parts the same.
It can been seen that the overly complex encoders (Capsule and ViT-based ones) turns out to be inferior to the CNN encoder.
Our assumption is that higher complexity increases the convergence difficulty and thus makes policy learning more challenging. 
\vspace{-7pt}
\begin{table*}[t!]
\addtolength{\tabcolsep}{-5pt}
\renewcommand{\arraystretch}{1.2}
\centering
\caption{\small
	Performance Comparison (STDEV$\leq$0.01) of different visual/audio encoder backbones.
}
\label{tab: fusion-encoder}
\vspace{0.1in}
\resizebox{1\linewidth}{!}{
\begin{tabular}{l|l|l|ccc|ccc|ccc|ccc|ccc|ccc} 
\hline
\multirow{4}{*}{Model} & \multirow{4}{*}{Fusion} & \multirow{4}{*}{Encoder}  & \multicolumn{9}{c|}{Replica}  & \multicolumn{9}{c}{Matterport3D}      \\ 
\cline{4-21}
    &       &      & \multicolumn{3}{c|}{Telephone} 
                   & \multicolumn{3}{c|}{Multiple heard} 
                   & \multicolumn{3}{c|}{Multiple unheard} 
                   & \multicolumn{3}{c|}{Telephone} 
                   & \multicolumn{3}{c|}{Multiple heard} 
                   & \multicolumn{3}{c}{Multiple unheard}  \\ 
\cline{4-21}
    &       &          & SPLT               & SSPLT             & SRT                
                       & SPLT               & SSPLT             & SRT                          
                       & SPLT               & SSPLT             & SRT                   
                       & SPLT               & SSPLT             & SRT                     
                       & SPLT               & SSPLT             & SRT                          
                       & SPLT               & SSPLT             & SRT                   \\ 
% \cline{3-20}
    &       &         & ($\uparrow$)       & ($\uparrow$)      & ($\uparrow$)                 
                       & ($\uparrow$)       & ($\uparrow$)      & ($\uparrow$)                     
                       & ($\uparrow$)       & ($\uparrow$)      & ($\uparrow$)               
                       & ($\uparrow$)       & ($\uparrow$)      & ($\uparrow$)                  
                       & ($\uparrow$)       & ($\uparrow$)      & ($\uparrow$)                       
                       & ($\uparrow$)       & ($\uparrow$)      & ($\uparrow$)                \\ 
\hline
 %--------------------------------------------------------------------------------------------------------------------
 FSAAVN   &  FSA   & CNN   & \textbf{0.541} & \textbf{0.635} & \textbf{0.925} & \textbf{0.438} & \textbf{0.541} & \textbf{0.812} & \textbf{0.182} & \textbf{0.316} & \textbf{0.358} & \textbf{0.520} & \textbf{0.585} & \textbf{0.832} & \textbf{0.438} & \textbf{0.496} & \textbf{0.844} & \textbf{0.207} & \textbf{0.299} & \textbf{0.391}    \\
 %--------------------------------------------------------------------------------------------------------------------
SoundSpaces        &  Concat   & CNN  & 0.531 & 0.604 & 0.892 & 0.354 & 0.462 & 0.764 & 0.152 & 0.255 & 0.317 & 0.454 & 0.511 & 0.797 & 0.431 & 0.475 & 0.818 & 0.180 & 0.254 & 0.350    \\
 %--------------------------------------------------------------------------------------------------------------------      
ViT-V          & Concat & ViT      & 0.521 & 0.584 & 0.871 & 0.329 & 0.415 & 0.713 &  0.138 & 0.233 & 0.304 & 0.412 & 0.465 & 0.797 & 0.012 & 0.188 & 0.027 & 0.013 & 0.170 & 0.020      \\
Capsule        & Concat & Capsule  & 0.426 & 0.503 & 0.810 & 0.262 & 0.372 & 0.580 &  0.154 & 0.278 & 0.330 & 0.317 & 0.382 & 0.742 & 0.246 & 0.302 & 0.623 & 0.178 & 0.255 & 0.445      \\
ViTScratch-V   & Concat & ViT      & 0.293 & 0.375 & 0.700 & 0.220 & 0.321 & 0.529 &  0.089 & 0.199 & 0.189 & 0.325 & 0.388 & 0.762 & 0.265 & 0.312 & 0.691 & 0.167 & 0.232 & 0.422      \\
 %--------------------------------------------------------------------------------------------------------------------                       
\hline
\end{tabular}
}
%\vspace{-5pt}
\end{table*}

\subsection{On dynamic modality importance}
\begin{figure*}[t!]
\centering
\includegraphics[width=\textwidth]{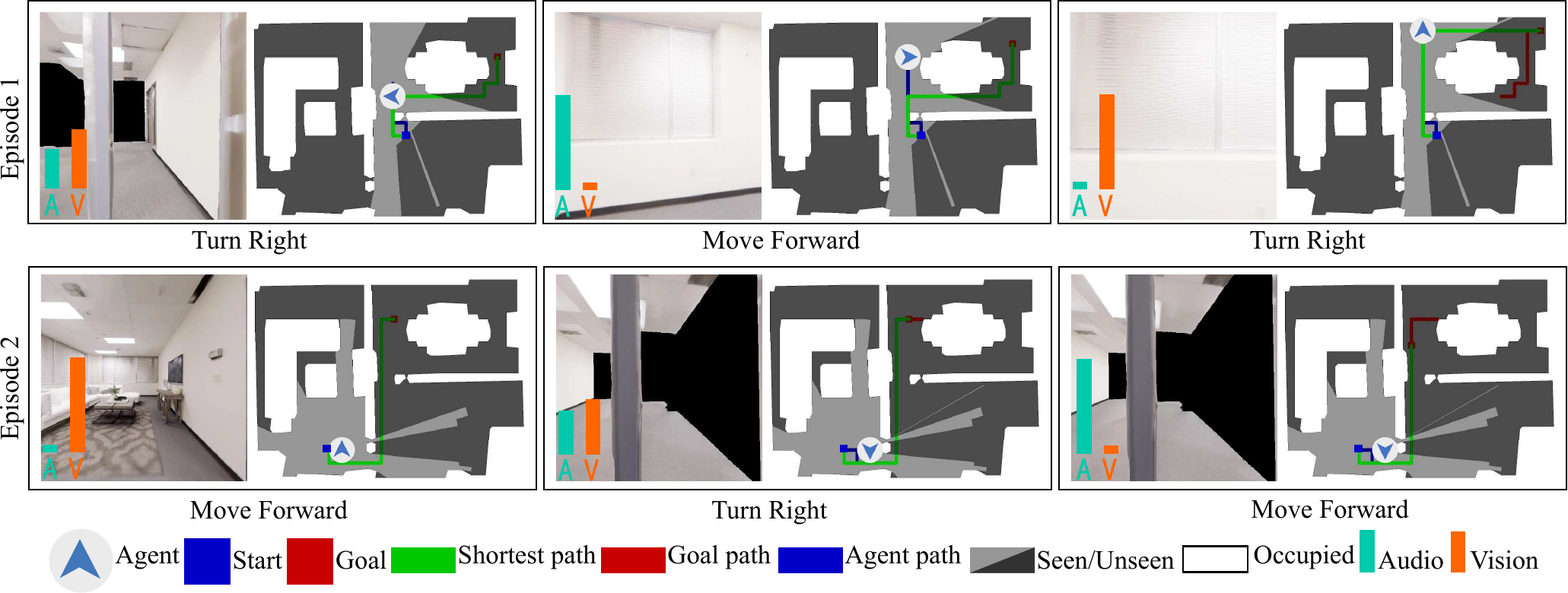}
\caption{\small 
	Dynamic visual and echo impact for two episodes. 
	Columns corresponds to three sampled time steps. The green and orange bars represent the importance of audio and vision, respectively. 
}
\label{fig: ImpactAudios}
%\vspace{-10pt}
\end{figure*}

Since the environmental context and target location (relative to the robot) changes all the time during navigation, we assume that the relative influence of audio and visual input on the robot's action can vary at different time points.
To quantify visual (audio) impact, we replace the visual (audio) input with random noise; 
and then we compute the visual (audio) impact score as the absolute difference (normalized) of the logarithmic action probabilities from the semi-corrupted model and the intact one.
Fig.\ref{fig: ImpactAudios} shows the impact scores (for two episodes in two rows) on the egocentric robot view at different time steps. 
We can see that FSAAVN dynamically re-weight the modalities (according to the current surroundings) while chasing after the moving audio target.

\subsection{Visualization of learned features and states}
\begin{figure*}[t!]
\centering
\includegraphics[width=\textwidth]{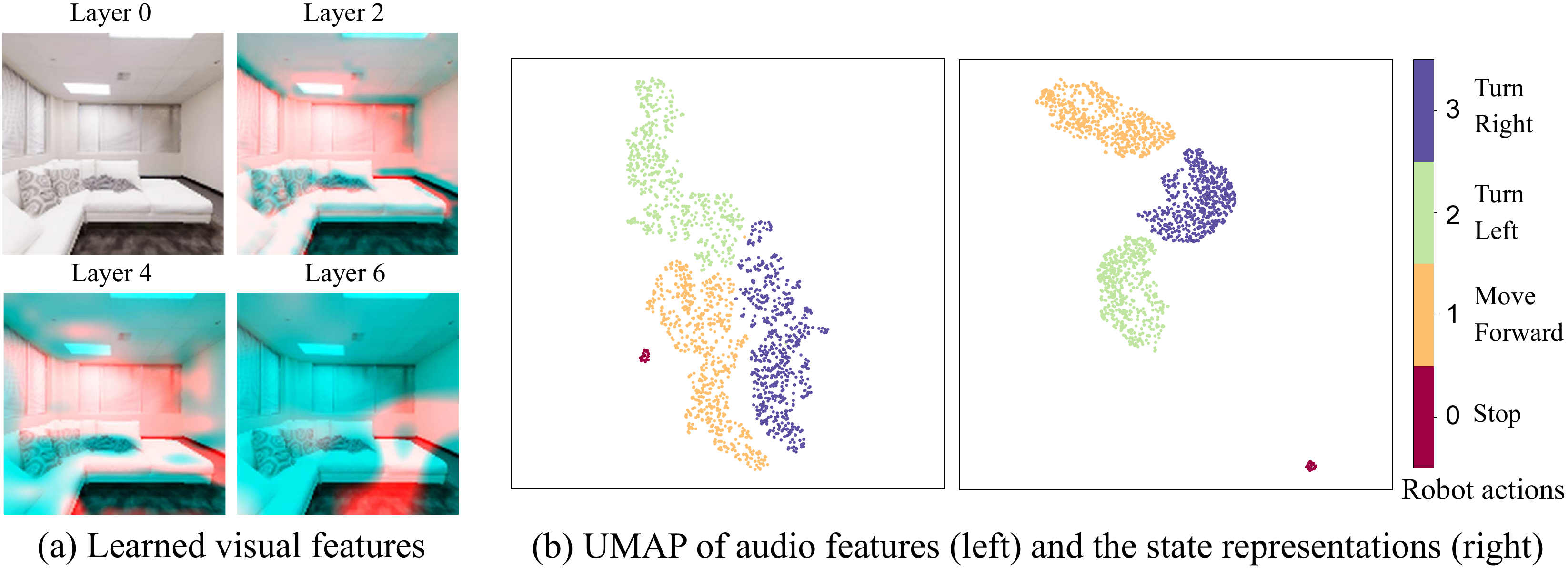}
\caption{\small Visualization of visual feature, audio feature, and state representations from Replica dataset.
}
\label{fig: fl}
%\vspace{-8pt}
\end{figure*}

In FSAAVN framework (cf.~Section~\ref{sec:approach}), the vision encoder generates visual feature $f_{I}(I_{t})$; the auditory encoder produces audio feature $f_{B}(B_{t})$; and the GRU transforms historical feature vectors into audio-visual state representations $s_t$.
The disengagement quality of these learned features and states is important to the downstream policy learning.
In Fig.\ref{fig: fl}(a), we examine the semantics of visual features by overlaying the output of the visual encoder (from different layers) over the RGB images.
It is easy to see that the visual encoder has learned to pay more attention to the area (in red color) where the robot can walk. 
This effect becomes more evident as the encoder becomes deeper.
It is more challenging to visualize the disengagement quality of audio features and the state representations, as a result, we choose to perform dimension reduction (to two dimensions) and clustering using UMAP~\cite{umap}.
The UMAP result is shown in Fig.\ref{fig: fl}(b) with a color coding representing the action selected by the robot.
Seen from Fig.\ref{fig: fl}(b), the learned audio features and state representations are naturally correlated with the robot's action selection.

\section{Conclusion} \label{sec:conclusion}
To realize a more effective (than the existing methods) audio-visual feature fusion strategy during audio-visual embodied navigation, we design a trainable Feature Self-Attention (FSA) module that determines the relative contribution of visual/audio modal in real-time in accordance with the ever-changing context.
We propose an end-to-end framework (FSAAVN: feature self-attention audio-visual navigation) incorporating FSA to train robots to catch up with a moving audio target.
FSAAVN is easy to train since it requires no extra aid like topology graph and sound semantics. 
Our comprehensive experiments validate the superior performance (both quantitatively and qualitatively) of FSAAVN in comparison with the state-of-the-arts.
We also carry out a set of thorough ablation studies on mainstream visual modalities, signal (visual/audio) encoders and audio-visual fusion strategies, providing useful insights for practitioners and researchers in this filed.

\section*{Acknowledgments}
% The acknowledgments are automatically included only in the final and preprint versions of the paper.
We thanks the insightful comments from the reviewers of BMVC 2022. This work is funded by Sino-German Collaborative Research Project {\it Crossmodal Learning} with identification number \verb|NSFC62061136001/DFG SFB/TRR169|.

\EnableMainend        %正文开关起作用的地方（结束）
%===============================================================================

%-------------------------------------------------------------------------
%-------------------------------------------------------------------------
%-------------------------------------------------------------------------
\EnableAppxstart      %附录部分起作用的地方 （开始）
% \fontsize{17}{60}\selectfont
\clearpage
% \title{ Supplementary Materials of Pay Self-Attention to Audio-Visual Navigation }
\title{\fontsize{16.5}{20}\selectfont Pay Self-Attention to Audio-Visual Navigation:\\ The Supplementary Material}
% \large{\fontsize{16.5}{20}\selectfont Pay Self-Attention to Audio-Visual Navigation:\\ The Supplementary Material}
% Enter the paper's authors in order
% \addauthor{Name}{email/homepage}{INSTITUTION_CODE}
\addauthor{Yinfeng Yu}{yyf17@mails.tsinghua.edu.cn}{1, 3, $*$}
\addauthor{Lele Cao}{lele.cao@eqtpartners.com}{1, 2, $*$}
% \addauthor{Fuchun Sun}{fcsun@mail.tsinghua.edu.cn}{1}
\addauthor{Fuchun Sun}{fcsun@mail.tsinghua.edu.cn}{1, \textrm{\Letter}}
\addauthor{Xiaohong Liu}{liu-xh17@mails.tsinghua.edu.cn}{1}
\addauthor{Liejun Wang}{wljxju@xju.edu.cn}{3}

% Enter the institutions
% \addinstitution{Name\\Address}
\addinstitution{
    Department of Computer Science \\
    and Technology, State Key Lab on \\
    Intelligent Technology and Systems, \\
    Tsinghua University, Beijing, China
}
\addinstitution{
    Motherbrain, EQT, Stockholm, Sweden
}
\addinstitution{
    College of Information Science \\
    and Engineering, Xinjiang University, \\
    Urumqi, China
}

% \runninghead{Student, Prof, Collaborator}{BMVC Author Guidelines}
\runninghead{Yu, Cao, Sun, Liu, Wang}{Pay Self-Attention to Audio-Visual Navigation}

%-------------------------------------------------------------------------
% Document starts here
\EnableMainstart 
\else
    \begin{document}
\EnableMainend

\maketitle
% 附录部分的公式从4开始，图从8开始，表从4开始
\setcounter{equation}{3}
\setcounter{figure}{7}
\setcounter{table}{3}
%--------------------------------------------------
%--------------------------------------------------
\vspace{-14pt}
%--------------------------------------------------
% \section*{\LARGE Content}
% \label{sec:appendix}
\appendix

\section*{Appendix}
\label{sec:appendix}

%--------------------------------------------------
This is the appendix (supplementary material) for the BMVC'22 paper titled ``Pay Self-Attention to Audio-Visual Navigation''. In that paper, we propose an end-to-end framework FSAAVN (feature self-attention audio-visual navigation) to learn chasing after a moving audio target using a context-aware audio-visual fusion strategy implemented as a Feature Self-Attention (FSA) module. The equations, tables and figures are numbered continuously in relation to the main paper. We provide additional details to the main paper concerning encoders, experiments and example navigation videos.

%--------------------------------------------------
% In this section, we provide additional details about:
%--------------------------------------------------
\ref{appx: encoders}:~~
\EnableMainstart 
    \EnableAppxstart
        %正文、附录的引用正常
        Specifications of various encoders. %(main paper ref.: \textsection~\ref{sec: experiments}).
    \EnableAppxend
\else
        %附录，硬写对正文的引用
        Specifications of various encoders. %(cf.~\textsection~5 in main paper).
\EnableMainend

%-----------------------------
% \ref{appx: relate-work-fig}:~~
% \EnableMainstart 
%     \EnableAppxstart
%         %正文、附录的引用正常
%         A supplementary figure for the related work (main paper ref.: \textsection~\ref{sec:relatedwork}).
%     \EnableAppxend
% \else
%         %附录，硬写对正文的引用
%         A supplementary figure for the related work (main paper ref.: \textsection~2).
% \EnableMainend
%-----------------------------

\ref{appx: exp}:~~
\EnableMainstart 
    \EnableAppxstart
        %正文、附录的引用正常
        Experimental details. %(main paper ref.: \textsection~\ref{sec: experiments}).
    \EnableAppxend
\else
        %附录，硬写对正文的引用
        Experimental details. % (main paper ref.: \textsection~5).
\EnableMainend

%-----------------------------
\hspace{0.2in}\ref{appx: metrics}:~~
\EnableMainstart 
    \EnableAppxstart
        %正文、附录的引用正常
        Evaluation metrics. % (main paper ref.: \textsection~\ref{sec: experiments}).
    \EnableAppxend
\else
        %附录，硬写对正文的引用
        Evaluation metrics. % (cf.~\textsection~5 in main paper).
\EnableMainend

%-----------------------------
%-----------------------------
\hspace{0.2in}\ref{appx: parameters}:~~
\EnableMainstart 
    \EnableAppxstart
        %正文、附录的引用正常
        Adopted experimental hyper-parameters. % (main paper ref.: \textsection~\ref{sec: experiments}).
    \EnableAppxend
\else
        %附录，硬写对正文的引用
        Adopted experimental hyper-parameters. %(main paper ref.: \textsection~5).
\EnableMainend

%-----------------------------
%-----------------------------
\hspace{0.2in}\ref{appx: compare-baselines}:~~
\EnableMainstart 
    \EnableAppxstart
        %正文、附录的引用正常
        Quantitative comparison of different algorithms on all datasets. %(main paper ref.: \textsection~\ref{sec: experiments}).
    \EnableAppxend
\else
        %附录，硬写对正文的引用
        Quantitative comparison of different algorithms on all datasets. %(main paper ref.: \textsection~5).
\EnableMainend

%-----------------------------
\hspace{0.2in}\ref{appx: fl-visual}:~~
\EnableMainstart 
    \EnableAppxstart
        %正文、附录的引用正常
        Visualization of the learned weights from visual encoder. %(main paper ref.: \textsection~\ref{sec: experiments}).
    \EnableAppxend
\else
        %附录，硬写对正文的引用
        % Visualization of weight data for each channel of each layer of the visual encoder (main paper ref.: \textsection~5).
        Visualization of the learned weights from visual encoder.
\EnableMainend

%-----------------------------
\hspace{0.2in}\ref{appx: traj-RGBD}:~~
\EnableMainstart 
    \EnableAppxstart
        %正文、附录的引用正常
        Navigation trajectories using RGBD input. %(main paper ref.: \textsection~\ref{sec: experiments}).
    \EnableAppxend
\else
        %附录，硬写对正文的引用
        % Navigation trajectories of algorithms FSAAVN and SoundSpaces on different datasets under RGBD input (main paper ref.: \textsection~5).
        Navigation trajectories using RGBD input.
\EnableMainend

%-----------------------------
\hspace{0.2in}\ref{appx: traj-depth}:~~
\EnableMainstart 
    \EnableAppxstart
        %正文、附录的引用正常
        Navigation trajectories using depth input. %(main paper ref.: \textsection~\ref{sec: experiments}).
    \EnableAppxend
\else
        %附录，硬写对正文的引用
        % Navigation trajectories of various algorithms on different datasets under Depth input (main paper ref.: \textsection~5).
        Navigation trajectories using depth input. %(main paper ref.: \textsection~5).
\EnableMainend

%-----------------------------

%-----------------------------
\ref{appendix: videos}:~~
\EnableMainstart 
    \EnableAppxstart
        %正文、附录的引用正常
        Example navigation video clips. %(main paper ref.: \textsection~\ref{sec: experiments}).
    \EnableAppxend
\else
        %附录，硬写对正文的引用
        % Video examples (main paper ref.: \textsection~5).
        Example navigation video clips. %(main paper ref.: \textsection~5).
\EnableMainend

%-----------------------------
% \ref{appendix: ethics}:~~
% Ethics statement.
%-------------------------------------------
%\clearpage
%----------------------------------------------
%
%----------------------------------------------------------------------------------

\section{Specifications of various encoders} \label{appx: encoders}
\paragraph{\textbf{CNN encoder.}}
Visual and auditory CNN encoders have separate weights but the same architecture of \verb|Conv8x8|, \verb|Conv4x4|, \verb|Conv3x3| and a linear layer; and ReLU activations are inserted between adjacent layers. 

\begin{figure*}[ht!]
	\centering
	\includegraphics[width=\textwidth]{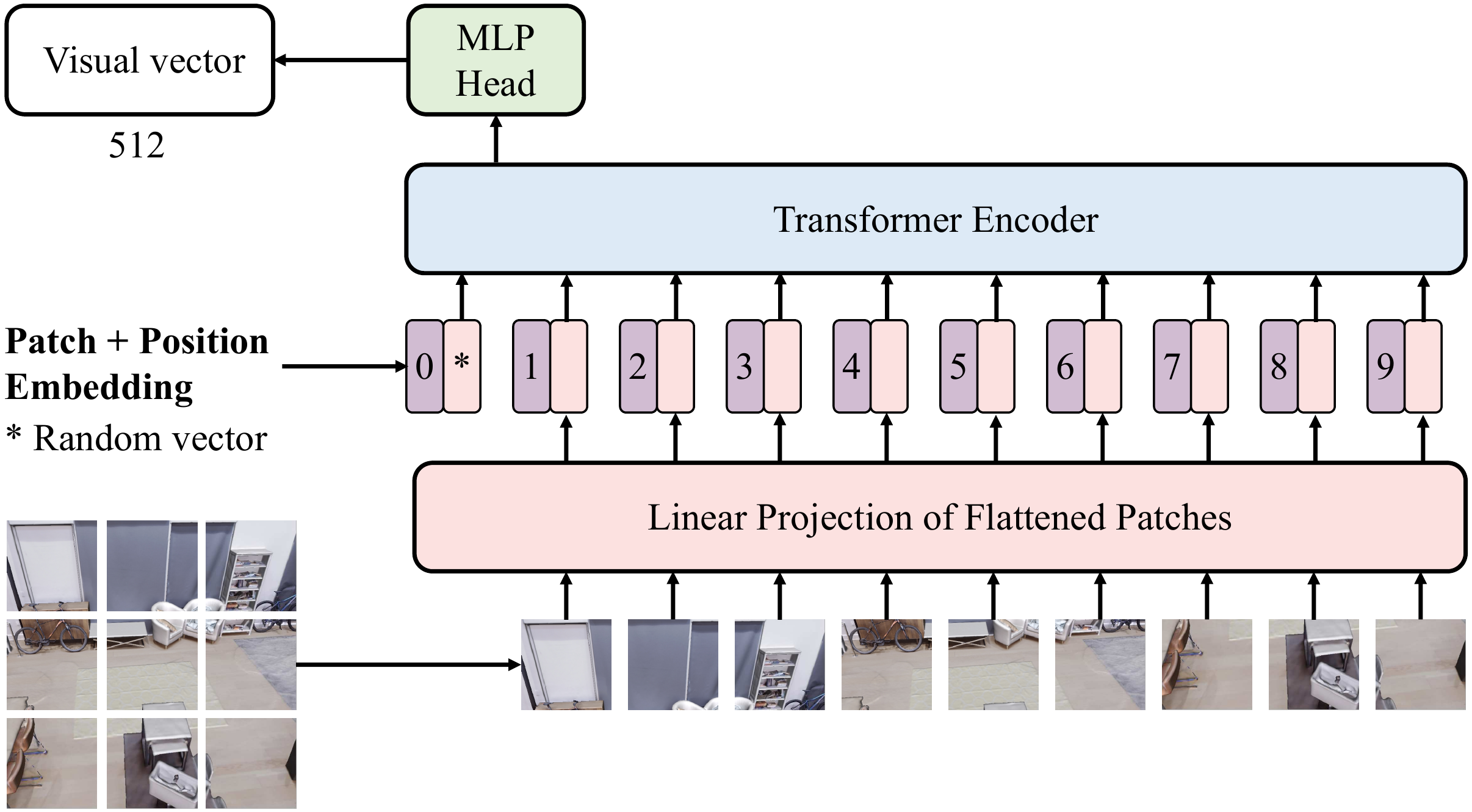}
	\caption{The implementation of Vision Transformer (ViT) encoder.
	}
	\label{fig: ViTEncoder}
 \vspace{10pt}
\end{figure*}

\begin{figure*}[ht!]
	\centering
	\includegraphics[width=\textwidth]{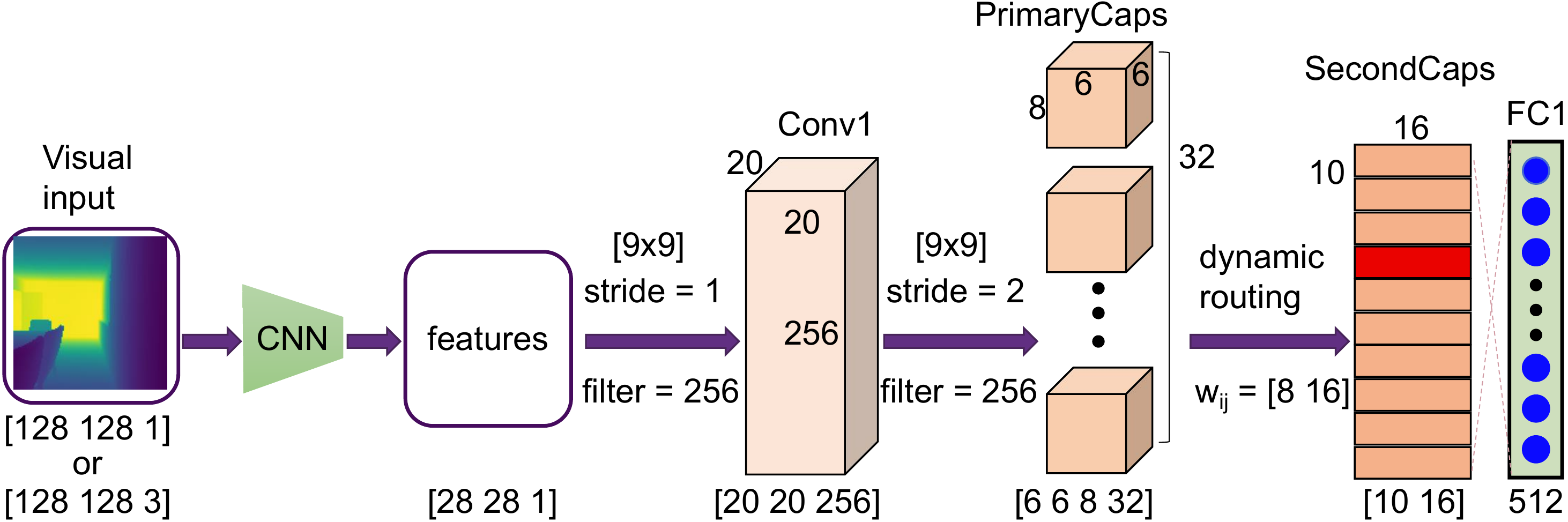}
	\caption{
	The implementation details of the capsule vision encoder.
	}
	\label{fig: capsuleEncoder}
\end{figure*}

\paragraph{\textbf{ViT encoder. }}
% original code from rwightman:
% https://github.com/rwightman/pytorch-image-models/blob/master/timm/models/vision_transformer.py
%# Flattern
% encoder_flatten = vit_tiny_patch16_224(num_classes = 1000, has_logits= False, device=device)
% #Modify input
% encoder_flatten.patch_embed = PatchEmbedForMultimodal(img_size=224, patch_size=16, in_c_dict=in_c_dict, embed_dim=192, norm_layer=None, flatten=True)
% #Modify the output so that it can output self._hidden_size = 512
% encoder_flatten.head = torch.nn.Linear(in_features=192, out_features=hidden_size, bias=True)
The backbone of ViT (Vision Transformer) is ``\verb|vit_tiny_patch16_224|'' defined in the \href{https://github.com/rwightman/pytorch-image-models/blob/master/timm/models/vision_transformer.py}{python file}. 
The input of the patch embed and the output head is modified so it can output a 512-dimensional visual vector.
The dimensions of the RGB image (\verb|128x128x3|), the depth image (\verb|128x128x1|) and the spectrum (\verb|65x69x2| for Replica and \verb|65x26x2| for Matterport3D) are different from the ones used in ViT encoders. 
We use resizing or transposed convolution to adapt the input sizes to \verb|224x224|.
The implementation details of the modified ViT encoder are shown in Fig.\ref{fig: ViTEncoder}.
%-----------------------------------------
%-----------------------------------------
\paragraph{\textbf{Capsule encoder.}}
Similar to the situation of ViT, the dimensions of the spectrum are also different from the ones used by capsule vision encoder. 
Therefore, resizing and transposed convolution are utilized to transform the input sizes to \verb|128x128|.
The implementation details of the capsule encoder are shown in Fig.\ref{fig: capsuleEncoder}.
% \clearpage
%-----------------------------------------

\section{Experiments} \label{appx: exp}
In this section, we will consecutively provide details about additional evaluation metrics, the values of hyper-parameters, quantitative comparison, visualization of visual encoder weights, and exemplary navigation trajectories.

\subsection{Evaluation metrics}\label{appx: metrics}
\EnableMainstart 
    \EnableAppxstart
        %正文、附录的引用正常
        The tracking metrics used in \textsection{\ref{sec: experiments}} of the main paper is elaborated in this section.
    \EnableAppxend
\else
    %附录，硬写对正文的引用
    In this section, the adopted evaluation metrics in addition to \textsection~5 of the main paper is elaborated below.
\EnableMainend

\begin{enumerate}[leftmargin=*]

    % \item \textbf{SPLT} (Success weighted by Path Length for Tracking):
    % \begin{equation}
    %     \mathrm{SPLT} = \frac{1}{N}\sum_{i=1}^{N}S_i\frac{l_i}{\mathrm{max}(p_i,l_i)},
    % \end{equation}
    % which weighs the successful episodes with the ratio of the shortest path $l_i$ to the executed path $p_i$, 
    % $S_{i}$ is the binary indicator of success(1)/failure(0) in the $i$-th episode $i$, and $N$ is the total number of episodes. 
    % When episode $i$ is finished with a success ($S_{i}=1$), the shortest path $l_i$ is extracted from the agent's start position to the goal's ending position.

    % \item \textbf{SSPLT} (Success weighted by Path Length of Tracking): 
    % \begin{equation}
    %     \mathrm{SSPLT} = \frac{1}{N}\sum_{i=1}^{N} \max (0,1- \frac{d^{a}_{i}}{d_{i}}) \frac{l_i}{\mathrm{max}(p_i,l_i)},
    % \end{equation}
    % where $l_i$ and $p_i$ is defined in the same way as when computing SPLT.
    % When computing SSPLT by the end of the $i$-th episode, $d^{a}_{i}$ is the robot's distance to the target, $d_{i}$ is the distance from the robot's start position to the target's final position.
    
    % \item \textbf{SRT} (Success Rate for Tracking):
    % \begin{equation}
    %     \mathrm{SRT}=\frac{1}{N}\sum^{N}_{i=1}S_{i}.
    % \end{equation}
    % SRT weighs the fraction of the completed episodes, where completion indicates that the agent catches the goal within the limit of 500 time steps and selects the stop action precisely at the target location. 
    
    \item \textbf{$R_{\text{mean}}$}: stands for average episode reward of the agent. 

    \item \textbf{NAT} (Number of Actions for Tracking): the number of executed actions by the agent.
    
    \item \textbf{SNAT} (Success weighted by Number of Actions of Tracking):
    \begin{equation}
        \mathrm{SNAT}=\frac{1}{N}\sum_{i=1}^{N}S_i\frac{l^a_i}{\mathrm{max}(p^a_i,l^a_i)},
    \end{equation}
    where $l^a_i$ is the number of actions taken for the shortest path from the agent's start position to the goal's final position when episode $i$ is finished, $p^a_i$ is the number of executed actions by the agent. Like SPLT and SSPLT in Equation~\eqref{eq:metrics}, this metric also captures the agent's efficiency in reaching the goal, yet SNAT accounts actions that do not lead to path changes, like rotation in place.
        
     \item \textbf{DTGT} (average Distance To Goal for Tracking):
     \begin{equation}
         \mathrm{DTGT}=\frac{1}{N}\sum^{N}_{i=1}d^{a}_{i},
     \end{equation}
     which weighs the agent's average distance to the goal when episodes are finished; $d^{a}_{i}$ is the agent's distance to the goal by the end of the $i$-th episode. 
        
     \item \textbf{NDTGT} (Normalized average Distance To Goal for Tracking):
     \begin{equation}
         \mathrm{NDTGT}=\frac{1}{N}\sum^{N}_{i=1}\frac{d^{a}_{i}}{d_{i}},
     \end{equation}
     where $d^{a}_{i}$ is the agent's distance to the goal when episode $i$ is finished,\,$d_{i}$ is the distance from the agent's start position to the goal's final position in the $i$-th episode. 
\end{enumerate}

\subsection{Adopted experimental hyper-parameters}\label{appx: parameters}
The values of hyper-parameters used in experimenting our model are specified in Table ~\ref{tab: parameters}.
\begin{table}[htb]
%\addtolength{\tabcolsep}{-3pt}
\renewcommand{\arraystretch}{0.9}
\centering
\caption{The adopted experimental hyper-parameters on Replica and Matterport3D datasets.}
\label{tab: parameters}
\vspace{0.15in}
\begin{tabular}{l|ll} 
\hline
Parameter                & Replica & Matterport3D    \\ 
\hline
RIR sampling rate      & 44100   & 16000   \\
clip parameter             & 0.1     & 0.1     \\
PPO epoch               & 4       & 4       \\
the number of mini batch         & 1       & 1       \\
value loss coefficient        & 0.5     & 0.5     \\
entropy coefficient            & 0.02    & 0.02    \\
learning rate                       & $2.5 \times 10^{-4}$  & $2.5 \times 10^{-4}$  \\
maximum gradient norm          & 0.5     & 0.5     \\
the number of steps               & 150     & 150     \\
use GAE (Generalized Advantage Estimation)                 & True    & True    \\
use linear learning rate  decay   & False   & False   \\
use linear clip decay & False   & False   \\
$\gamma$                    & 0.99    & 0.99    \\
$\tau$                      & 0.95    & 0.95    \\
$\beta$                      & 0.01    & 0.01    \\
reward window size     & 50      & 50      \\
success reward          & 10.0    & 10.0    \\
slack reward            & -0.01   & -0.01   \\
distance reward scale  & 1.0     & 1.0     \\
hidden size             & 512     & 512     \\
the type of optimizer               & Adam     & Adam   \\
the number of processes     & 5         & 10     \\
the number of updates       & 40000      & 60000 \\
the probability of sound source moving position & 0.3 & 0.3 \\
\hline
\end{tabular}
\end{table}

\subsection{Quantitative comparison of different algorithms on all datasets} \label{appx: compare-baselines}
The quantitative benchmarking results on Replica and Matterport3D datasets are shown in Table~\ref{appx: comparison-replica} and \ref{appx: comparison-mp3d}, respectively.
\begin{table*}[ht!]
\addtolength{\tabcolsep}{-3pt}
\renewcommand{\arraystretch}{1.2}
\centering
\caption{
Quantitative comparison of different algorithms on Replica dataset. All results are averaged over 5 test runs.
}
\label{appx: comparison-replica}
\vspace{0.12in}
\resizebox{1.0\linewidth}{!}{
\begin{tabular}{lll|cccccccc}
\hline
Method      & Vision  & Audio   & SPLT ($\uparrow$)       & SSPLT ($\uparrow$)      & SRT  ($\uparrow$)       & $R_{mean}$ ($\uparrow$)   & NAT ($\downarrow$)        & SNAT ($\uparrow$)       & DTGT ($\downarrow$)      & NDTGT ($\downarrow$)          \\
\hline
%--------------------------------------------------------
FSAAVN    & Depth  & Telephone       & \textbf{0.541±0.004} & \textbf{0.635±0.003} & \textbf{0.925±0.005} & \textbf{17.2±0.1} & \textbf{57.2±0.5}  & \textbf{0.316±0.004} & \textbf{0.01±0.01} & \textbf{9.0±0.1}    \\
SoundSpaces        & Depth   & Telephone      & 0.531±0.003 & 0.604±0.001 & 0.892±0.003 & 16.9±0.1 & 60.7±0.2  & 0.305±0.004 & 0.03±0.01 & 70.0±1.5    \\
SoundSpaces-EMul     & Depth  & Telephone       & 0.493±0.005 & 0.597±0.001 & 0.861±0.004 & 16.6±0.1 & 61.7±0.2  & 0.280±0.003 & 0.16±0.01 & 100.1±0.4   \\
SoundSpaces-EM      & Depth   & Telephone      & 0.487±0.002 & 0.592±0.002 & 0.816±0.003 & 16.3±0.1 & 62.7±1.3  & 0.272±0.001 & 0.19±0.01 & 130.1±0.7   \\
CMHM       & Depth   & Telephone      & 0.335±0.005 & 0.338±0.004 & 0.791±0.020 & 16.0±0.1 & 87.3±1.6  & 0.234±0.004 & 0.24±0.01 & 230.1±1.1      \\
AV-WaN        & Depth  & Telephone       & 0.218±0.003 & 0.224±0.004 & 0.764±0.002 & 14.9±0.2 & 182.2±4.9 & 0.167±0.004 & 0.43±0.06 & 220.1±1.1   \\

\hline
%--------------------------------------------------------
FSAAVN    & Depth & Multiple heard     & \textbf{0.438±0.001} & \textbf{0.541±0.001} & \textbf{0.812±0.002} & \textbf{16.8±0.1} & \textbf{75.0±0.5}  & \textbf{0.229±0.001} & \textbf{0.14±0.01} & \textbf{10.1±0.2}     \\
SoundSpaces-EM      & Depth & Multiple heard     & 0.435±0.001 & 0.531±0.001 & 0.796±0.003 & 16.6±0.1 & 81.6±1.4  & 0.224±0.001 & 0.17±0.01 & 80.1±0.4     \\
SoundSpaces-EMul      & Depth & Multiple heard     & 0.430±0.001 & 0.522±0.001 & 0.770±0.001 & 16.3±0.2 & 82.8±1.2  & 0.216±0.001 & 0.20±0.01 & 100.1±0.6     \\
SoundSpaces        & Depth & Multiple heard     & 0.354±0.003 & 0.462±0.002 & 0.764±0.005 & 15.7±0.1 & 83.0±1.7  & 0.188±0.003 & 0.22±0.01 & 150.1±0.7    \\
CMHM       & Depth & Multiple heard     & 0.259±0.003 & 0.302±0.002 & 0.692±0.005 & 15.4±0.1 & 127.2±0.3 & 0.179±0.004 & 0.37±0.02 & 460.1±2.2     \\
AV-WaN  & Depth & Multiple heard     & 0.220±0.002 & 0.271±0.004 & 0.533±0.004 & 11.6±0.1 & 215.6±3.2 & 0.011±0.003 & 1.33±0.07 & 690.4±3.6     \\

\hline
%--------------------------------------------------------
FSAAVN    & Depth   &  Multiple unheard     & \textbf{0.182±0.002} & \textbf{0.316±0.003} & \textbf{0.358±0.003} & \textbf{8.0±0.1}  & \textbf{227.0±1.9} & \textbf{0.084±0.001}  & \textbf{1.79±0.01} & \textbf{790.5±3.5}   \\
SoundSpaces-EMul      & Depth   &  Multiple unheard     & 0.168±0.001 & 0.304±0.001 & 0.326±0.001 & 7.9±0.1  & 236.0±2.0 & 0.082±0.001  & 1.90±0.01 & 813.4±4.1     \\
SoundSpaces-EM      & Depth   &  Multiple unheard     & 0.154±0.001 & 0.258±0.001 & 0.319±0.001 & 7.6±0.1  & 250.5±1.7 & 0.080±0.002  & 2.05±0.01 & 839.1±4.4     \\
SoundSpaces        & Depth   &  Multiple unheard     & 0.152±0.001 & 0.255±0.001 & 0.317±0.001 & 7.4±0.1  & 256.9±2.5 & 0.078±0.002  & 2.13±0.03 & 860.7±4.5   \\
CMHM       & Depth   &  Multiple unheard     & 0.121±0.003 & 0.202±0.004 & 0.314±0.001 & 6.9±0.1  & 265.1±2.2 & 0.046±0.002  & 2.21±0.01 & 1460.6±7.7  \\
AV-WaN  & Depth   &  Multiple unheard     & 0.010±0.004 & 0.189±0.004 & 0.233±0.002 & 5.6±0.2  & 304.3±1.9 & 0.031±0.001  & 2.63±0.07 & 1890.5±9.1   \\

\hline
%--------------------------------------------------------
FSAAVN    & RGBD  & Telephone        & \textbf{0.532±0.002} & \textbf{0.611±0.003} & \textbf{0.837±0.001} & \textbf{17.2±0.1} & \textbf{63.7±0.3}  & \textbf{0.285±0.001}  & \textbf{0.14±0.01} & \textbf{190.1±0.9}  \\
SoundSpaces        & RGBD   & Telephone       & 0.527±0.004 & 0.605±0.003 & 0.835±0.001 & 17.1±0.1 & 66.4±0.2  & 0.282±0.001  & 0.15±0.01 & 450.1±2.1   \\
\hline
%--------------------------------------------------------
FSAAVN    & RGBD & Multiple heard      & \textbf{0.402±0.001} & \textbf{0.485±0.001} & \textbf{0.792±0.003} & \textbf{16.3±0.1} & \textbf{92.2±0.2}  & \textbf{0.189±0.001}  & \textbf{0.23±0.01} & \textbf{290.1±1.4}  \\
AVN        & RGBD & Multiple heard      & 0.393±0.002 & 0.475±0.001 & 0.756±0.003 & 16.1±0.1 & 99.5±0.1  & 0.187±0.001  & 0.24±0.01 & 330.1±1.6   \\
\hline
%--------------------------------------------------------
FSAAVN   & RGBD & Multiple unheard   & \textbf{0.185±0.001} & \textbf{0.285±0.001} & \textbf{0.349±0.001} & \textbf{7.4±0.1}  & \textbf{271.1±0.2} & \textbf{0.085±0.004}  & \textbf{2.08±0.01} & \textbf{990.5±4.7}  \\
SoundSpaces       & RGBD & Multiple unheard   & 0.182±0.001 & 0.276±0.001 & 0.339±0.002 & 7.0±0.1  & 295.0±0.1 & 0.030±0.001  & 2.11±0.01 & 2250.6±11.1   \\
\hline
%--------------------------------------------------------
FSAAVN   & RGB   & Telephone         & \textbf{0.530±0.003} & \textbf{0.601±0.003} & \textbf{0.872±0.003} & \textbf{17.2±0.1} & \textbf{69.5±0.3}  & \textbf{0.273±0.002}  & \textbf{0.14±0.01} & \textbf{130.1±0.7}       \\
SoundSpaces       & RGB  & Telephone          & 0.522±0.003 & 0.593±0.004 & 0.829±0.004 & 17.0±0.1 & 72.4±0.2  & 0.262±0.001  & 0.22±0.02 & 375.1±1.8        \\
\hline
%--------------------------------------------------------
FSAAVN   & RGB & Multiple heard        & \textbf{0.413±0.002} & \textbf{0.500±0.002} & \textbf{0.767±0.002} & \textbf{15.9±0.1} & \textbf{94.3±0.3}  & \textbf{0.190±0.003}  & \textbf{0.28±0.01} & \textbf{140.1±0.7}        \\
SoundSpaces       & RGB & Multiple heard        & 0.386±0.004 & 0.477±0.003 & 0.741±0.003 & 15.6±0.1 & 97.3±0.2  & 0.184±0.002  & 0.30±0.01 & 610.1±3.7        \\
\hline
%--------------------------------------------------------
FSAAVN   & RGB & Multiple unheard    & \textbf{0.166±0.002} & \textbf{0.295±0.005} & \textbf{0.305±0.002} & \textbf{7.1±0.2}  & \textbf{255.1±0.2} & \textbf{0.073±0.003}  & \textbf{2.10±0.02} & \textbf{1230.6±6.8}       \\
SoundSpaces       & RGB & Multiple unheard    & 0.140±0.003 & 0.260±0.004 & 0.267±0.004 & 6.3±0.2  & 285.2±0.3 & 0.066±0.004  & 2.17±0.01 & 1270.6±6.4       \\
\hline
%--------------------------------------------------------
FSAAVN   & Blind    & Telephone      & \textbf{0.470±0.001} & \textbf{0.544±0.001} & \textbf{0.833±0.001} & \textbf{17.1±0.1} & \textbf{74.3±0.1}  & \textbf{0.225±0.001}  & \textbf{0.14±0.01} & \textbf{183.4±0.9}   \\
SoundSpaces       & Blind  & Telephone        & 0.472±0.001 & 0.545±0.001 & 0.839±0.001 & 17.1±0.1 & 73.3±0.1  & 0.233±0.001  & 0.13±0.01 & 50.0±0.3   \\
\hline
%--------------------------------------------------------
FSAAVN   & Blind & Multiple heard      & \textbf{0.328±0.005} & \textbf{0.425±0.001} & \textbf{0.703±0.004} & \textbf{15.2±0.1} & \textbf{99.9±0.1}  & \textbf{0.141±0.001}  & \textbf{0.37±0.01} & \textbf{200.1±0.9}  \\
SoundSpaces       & Blind & Multiple heard      & 0.334±0.003 & 0.425±0.001 & 0.725±0.005 & 15.3±0.1 & 99.5±0.1  & 0.143±0.001  & 0.32±0.01 & 183.5±0.8    \\
\hline
%--------------------------------------------------------
FSAAVN   & Blind & Multiple unheard  & \textbf{0.141±0.001} & \textbf{0.229±0.001} & \textbf{0.294±0.004} & \textbf{6.3±0.2}  & \textbf{290.8±0.3} & \textbf{0.060±0.001}  & \textbf{2.43±0.03} & \textbf{1450.6±7.6}    \\
SoundSpaces       & Blind & Multiple unheard  & 0.142±0.001 & 0.229±0.001 & 0.331±0.005 & 6.7±0.0  & 276.6±0.1 & 0.062±0.001  & 2.35±0.01 & 350.6±1.7   \\
\hline
%----------------------------------------------------------------------------------------
ViT-V         & Depth  & Telephone    & 0.521±0.003 & 0.584±0.001 & 0.871±0.002 & 17.5±0.1 & 69.1±1.8  & 0.265±0.004 & 0.12±0.02 & 70.0±0.8    \\
Capsule-V     & Depth  & Telephone    & 0.467±0.002 & 0.539±0.002 & 0.835±0.002 & 17.3±0.1 & 75.8±1.9  & 0.218±0.003 & 0.14±0.01 & 130.1±0.6  \\
Capsule       & Depth  & Telephone    & 0.426±0.001 & 0.503±0.003 & 0.810±0.003 & 17.0±0.1 & 82.7±2.1  & 0.203±0.004 & 0.16±0.01 & 240.1±1.2   \\
ViTScratch-V  & Depth  & Telephone    & 0.293±0.001 & 0.375±0.005 & 0.700±0.005 & 15.3±0.1 & 123.0±2.3 & 0.122±0.005 & 0.42±0.03 & 530.1±2.4   \\
\hline
%----------------------------------------------------------------------------------------
ViT-V        & Depth & Multiple heard   & 0.329±0.001 & 0.415±0.001 & 0.713±0.001 & 15.3±0.1 & 97.4±0.8  & 0.152±0.002 & 0.35±0.01 & 210.1±1.2  \\ 
Capsule-V    & Depth & Multiple heard   & 0.320±0.001 & 0.413±0.001 & 0.695±0.001 & 15.0±0.1 & 105.5±0.8 & 0.144±0.002 & 0.40±0.01 & 379.1±1.9    \\
Capsule      & Depth & Multiple heard   & 0.262±0.002 & 0.372±0.001 & 0.580±0.011 & 13.1±0.2 & 106.9±0.9 & 0.114±0.001 & 0.95±0.01 & 382.2±1.9   \\
ViTScratch-V & Depth & Multiple heard   & 0.220±0.001 & 0.321±0.002 & 0.529±0.021 & 12.6±0.2 & 125.2±2.0 & 0.097±0.001 & 0.98±0.01 & 640.3±3.4   \\
\hline
%----------------------------------------------------------------------------------------
Capsule      & Depth   &  Multiple unheard   & 0.154±0.001 & 0.278±0.003 & 0.330±0.004 & 7.7±0.1  & 208.0±1.6 & 0.068±0.001 & 2.41±0.01 & 1000.6±5.0  \\
ViT-V        & Depth   &  Multiple unheard   & 0.138±0.001 & 0.233±0.004 & 0.304±0.001 & 6.5±0.1  & 273.4±1.2 & 0.064±0.001 & 2.42±0.01 & 1050.8±5.1   \\
Capsule-V    & Depth   &  Multiple unheard   & 0.133±0.001 & 0.207±0.002 & 0.296±0.002 & 6.0±0.1  & 277.3±1.0 & 0.059±0.001 & 2.46±0.01 & 1140.6±5.9  \\
ViTScratch-V & Depth   &  Multiple unheard   & 0.089±0.001 & 0.199±0.001 & 0.189±0.003 & 4.4±0.1  & 310.6±1.1 & 0.040±0.003 & 3.30±0.01 & 2230.6±10.5  \\
\hline
Capsule-V-EM   & Depth & Telephone   & 0.527±0.002 & 0.602±0.003 & 0.852±0.003 & 17.5±0.1 & 64.3±0.3  & 0.273±0.001 & 0.12±0.01 & 110.1±0.6  \\
ViT-V-EMul     & Depth & Telephone  & 0.481±0.002 & 0.551±0.001 & 0.843±0.003 & 17.2±0.1 & 67.3±0.4  & 0.269±0.001 & 0.15±0.01 & 150.0±0.7   \\
Capsule-V-EMul & Depth & Telephone  & 0.475±0.003 & 0.540±0.001 & 0.837±0.005 & 17.0±0.1 & 71.7±0.4  & 0.240±0.005 & 0.17±0.01 & 220.0±1.1   \\
ViT            & Depth & Telephone  & 0.449±0.005 & 0.520±0.005 & 0.815±0.003 & 16.5±0.1 & 77.1±0.5  & 0.222±0.004 & 0.22±0.03 & 260.1±1.3    \\
ViT-V-EM       & Depth & Telephone  & 0.421±0.004 & 0.517±0.004 & 0.766±0.002 & 16.3±0.1 & 82.7±0.7  & 0.208±0.004 & 0.30±0.02 & 390.1±1.9  \\
ViTScratch     & Depth & Telephone  & 0.162±0.005 & 0.277±0.005 & 0.423±0.004 & 9.5±0.2  & 129.0±2.0 & 0.070±0.003 & 1.39±0.05 & 920.4±4.7  \\
\hline
\end{tabular}
}
\end{table*}
%-----------------------------------------
\clearpage
\begin{table*}[ht!]
\addtolength{\tabcolsep}{-3pt}
\renewcommand{\arraystretch}{1.2}
\centering
\caption{
Quantitative comparison of different algorithms on Matterport3D dataset. All results are averaged over 5 test runs.
}
\label{appx: comparison-mp3d}
\vspace{0.15in}
\resizebox{1.0\linewidth}{!}{
\begin{tabular}{lll|cccccccc}
\hline
Method      & Vision  & Audio   & SPLT ($\uparrow$)       & SSPLT ($\uparrow$)      & SRT  ($\uparrow$)       & $R_{mean}$ ($\uparrow$)   & NAT ($\downarrow$)        & SNAT ($\uparrow$)       & DTGT ($\downarrow$)      & NDTGT ($\downarrow$)          \\
\hline
FSAAVN           & Depth  & Telephone        & \textbf{0.520±0.002} & \textbf{0.585±0.002} & \textbf{0.832±0.002} & \textbf{31.3±0.1} & \textbf{112.1±1.1} & \textbf{0.308±0.001} & \textbf{2.24±0.02}  & \textbf{20.1±0.1}  \\
SoundSpaces-EM   & Depth & Telephone          & 0.481±0.001 & 0.543±0.002 & 0.817±0.001 & 30.7±0.1 & 119.8±1.2 & 0.288±0.002 & 2.56±0.04  & 120.1±0.6 \\
SoundSpaces-EMul & Depth & Telephone          & 0.457±0.001 & 0.523±0.001 & 0.801±0.001 & 30.3±0.1 & 122.8±1.3 & 0.276±0.001 & 3.02±0.02  & 220.1±1.1   \\
SoundSpaces      & Depth & Telephone          & 0.454±0.001 & 0.511±0.002 & 0.797±0.001 & 30.0±0.1 & 126.4±1.1 & 0.270±0.001 & 3.92±0.03  & 300.1±1.5      \\   
CMHM       & Depth       & Telephone    & 0.114±0.001 & 0.125±0.001 & 0.606±0.001 & 20.1±0.1 & 298.4±1.2 & 0.076±0.002 & 8.20±0.04  & 1100.4±5.0  \\
AV-WaN  & Depth          & Telephone  & 0.111±0.001 & 0.114±0.002 & 0.409±0.001 & 16.9±0.1 & 333.6±1.5 & 0.047±0.001 & 10.53±0.04 & 1440.6±7.2   \\

\hline
FSAAVN           & Depth   &  Multiple heard        & \textbf{0.438±0.001} & \textbf{0.496±0.003} & \textbf{0.844±0.002} & \textbf{31.0±0.1} & \textbf{123.3±1.1} & \textbf{0.268±0.001} & \textbf{2.23±0.01}  & \textbf{100.1±0.5}    \\
SoundSpaces-EM   & Depth   &  Multiple heard        & 0.435±0.001 & 0.492±0.002 & 0.832±0.001 & 30.6±0.1 & 126.1±1.2 & 0.254±0.002 & 2.35±0.02  & 140.1±0.7  \\
SoundSpaces-EMul & Depth   &  Multiple heard        & 0.433±0.001 & 0.481±0.002 & 0.821±0.001 & 30.2±0.1 & 137.0±0.3 & 0.235±0.001 & 2.90±0.05  & 240.1±1.2    \\
SoundSpaces      & Depth   &  Multiple heard        & 0.431±0.001 & 0.475±0.002 & 0.818±0.001 & 29.6±0.1 & 144.5±0.8 & 0.223±0.002 & 3.17±0.01  & 650.1±3.2    \\
CMHM             & Depth   &  Multiple heard        & 0.086±0.002 & 0.099±0.001 & 0.528±0.002 & 17.4±0.2 & 323.6±1.8 & 0.056±0.001 & 9.80±0.03  & 1360.5±6.8   \\
AV-WaN           & Depth   &  Multiple heard        & 0.012±0.001 & 0.034±0.001 & 0.093±0.001 & 3.0±0.1  & 439.9±1.2 & 0.009±0.001 & 18.76±0.03 & 2901.1±15.3  \\

\hline
FSAAVN           & Depth   &  Multiple unheard       & \textbf{0.207±0.002} & \textbf{0.299±0.001} & \textbf{0.391±0.015} & \textbf{18.8±0.1} & \textbf{287.8±0.2} & \textbf{0.113±0.001} & \textbf{7.91±0.02}  & \textbf{200.5±1.2}   \\
SoundSpaces-EM   & Depth   &  Multiple unheard        & 0.183±0.001 & 0.266±0.001 & 0.375±0.001 & 18.6±0.1 & 291.7±0.7 & 0.108±0.001 & 8.16±0.02  & 400.5±2.2    \\
SoundSpaces-EMul & Depth   &  Multiple unheard        & 0.182±0.001 & 0.258±0.001 & 0.355±0.001 & 17.7±0.1 & 297.8±0.7 & 0.104±0.001 & 8.23±0.03  & 440.7±2.9      \\
SoundSpaces      & Depth   &  Multiple unheard       & 0.180±0.001 & 0.254±0.002 & 0.350±0.001 & 17.3±0.1 & 313.6±0.8 & 0.097±0.001 & 8.56±0.02  & 1300.5±6.8   \\
CMHM             & Depth   &  Multiple unheard       & 0.052±0.002 & 0.085±0.002 & 0.267±0.002 & 11.3±0.2 & 317.1±0.9 & 0.034±0.002 & 13.68±0.15 & 1461.1±7.0    \\
AV-WaN           & Depth   &  Multiple unheard       & 0.010±0.002 & 0.043±0.003 & 0.057±0.002 & 4.0±0.1  & 385.0±0.8 & 0.007±0.001 & 18.06±0.49 & 2600.4±10.0  \\

\hline
FSAAVN       & RGBD  & Telephone           & \textbf{0.454±0.003} & \textbf{0.510±0.001} & \textbf{0.834±0.004} & \textbf{30.9±0.1} & \textbf{120.1±0.7} & \textbf{0.281±0.005} & \textbf{2.45±0.02}  & \textbf{40.1±0.2}   \\
SoundSpaces  & RGBD  & Telephone          & 0.435±0.005 & 0.502±0.002 & 0.798±0.002 & 30.6±0.1 & 129.5±1.0 & 0.240±0.003 & 2.51±0.01  & 280.1±1.2   \\

\hline
FSAAVN       & RGBD & Multiple heard          & \textbf{0.440±0.004} & \textbf{0.492±0.005} & \textbf{0.827±0.003} & \textbf{30.3±0.1} & \textbf{125.0±0.9} & \textbf{0.275±0.004} & \textbf{2.70±0.04}  & \textbf{215.1±1.1}  \\
SoundSpaces  & RGBD & Multiple heard          & 0.412±0.003 & 0.469±0.004 & 0.809±0.005 & 30.0±0.1 & 135.6±0.7 & 0.243±0.003 & 2.80±0.05  & 220.1±1.1    \\

\hline
FSAAVN       & RGBD & Multiple unheard      & \textbf{0.191±0.001} & \textbf{0.281±0.001} & \textbf{0.373±0.001} & \textbf{18.8±0.1} & \textbf{306.0±0.1} & \textbf{0.119±0.001} & \textbf{7.78±0.03}  & \textbf{1600.5±8.8}  \\
SoundSpaces  & RGBD & Multiple unheard      & 0.186±0.001 & 0.277±0.001 & 0.369±0.001 & 18.5±0.1 & 314.5±0.2 & 0.108±0.001 & 7.94±0.04  & 2400.5±12.9  \\

\hline
FSAAVN      & RGB  & Telephone             & \textbf{0.449±0.002} & \textbf{0.505±0.003} & \textbf{0.820±0.002} & \textbf{30.5±0.1} & \textbf{143.0±0.3} & \textbf{0.212±0.001} & \textbf{2.57±0.01}  & \textbf{60.1±0.4}      \\
SoundSpaces & RGB  & Telephone             & 0.397±0.003 & 0.451±0.003 & 0.815±0.001 & 30.2±0.1 & 150.6±0.1 & 0.195±0.001 & 2.58±0.01  & 80.1±0.7      \\
\hline
FSAAVN      & RGB  & Multiple heard          & \textbf{0.393±0.003} & \textbf{0.453±0.002} & \textbf{0.781±0.001} & \textbf{29.6±0.1} & \textbf{158.3±0.2} & \textbf{0.189±0.004} & \textbf{3.12±0.02}  & \textbf{300.1±1.5}  \\
SoundSpaces & RGB  & Multiple heard          & 0.371±0.002 & 0.429±0.004 & 0.772±0.002 & 28.9±0.1 & 166.5±0.3 & 0.180±0.003 & 3.57±0.03  & 340.1±1.7    \\
\hline
FSAAVN      & RGB  & Multiple unheard      & \textbf{0.196±0.001} & \textbf{0.270±0.001} & \textbf{0.417±0.002} & \textbf{18.4±0.2} & \textbf{299.1±0.6} & \textbf{0.099±0.003} & \textbf{8.60±0.05}  & \textbf{740.5±2.3}  \\
SoundSpaces & RGB  & Multiple unheard      & 0.193±0.001 & 0.269±0.001 & 0.375±0.004 & 17.2±0.1 & 327.8±0.4 & 0.093±0.002 & 8.93±0.04  & 1000.5±5.6     \\
\hline
FSAAVN      & Blind  & Telephone        & 0.369±0.001 & 0.424±0.003 & 0.787±0.001 & 29.5±0.1 & 171.1±0.5 & 0.158±0.004 & 3.04±0.01  & 340.1±1.7     \\
SoundSpaces & Blind  & Telephone        & 0.385±0.002 & 0.443±0.004 & 0.790±0.001 & 29.7±0.1 & 167.4±1.1 & 0.167±0.004 & 2.94±0.01  & 40.1±0.4      \\

\hline
FSAAVN      & Blind  & Multiple heard      & 0.339±0.002 & 0.387±0.001 & 0.766±0.006 & 28.6±0.2 & 179.8±5.0 & 0.158±0.005 & 3.42±0.02  & 220.2±1.3   \\
SoundSpaces & Blind  & Multiple heard      & 0.319±0.001 & 0.372±0.001 & 0.724±0.001 & 27.5±0.1 & 195.1±2.5 & 0.133±0.003 & 4.09±0.03  & 240.2±1.5    \\

\hline
FSAAVN      & Blind  & Multiple unheard  & 0.162±0.001 & 0.241±0.002 & 0.356±0.001 & 17.0±0.1 & 314.8±1.3 & 0.069±0.001 & 9.47±0.05  & 1020.5±5.8  \\
SoundSpaces & Blind  & Multiple unheard  & 0.163±0.001 & 0.257±0.001 & 0.370±0.001 & 17.6±0.1 & 295.4±1.1 & 0.074±0.001 & 8.70±0.05  & 800.5±4.9  \\

\hline
ViT-V           & Depth  & Telephone  & 0.412±0.002 & 0.465±0.004 & 0.797±0.003 & 29.5±0.1 & 148.1±1.0 & 0.207±0.001 & 3.08±0.10  & 40.1±0.3    \\
ViTScratch-V    & Depth  & Telephone  & 0.325±0.003 & 0.388±0.001 & 0.762±0.002 & 29.0±0.1 & 182.2±1.1 & 0.142±0.001 & 3.32±0.10  & 80.1±0.7     \\
Capsule         & Depth  & Telephone  & 0.317±0.002 & 0.382±0.001 & 0.742±0.004 & 28.6±0.1 & 185.5±1.1 & 0.136±0.001 & 3.59±0.10  & 260.1±1.2    \\
\hline
%-----------------------------------------------------------------------
ViTScratch-V   & Depth   &  Multiple heard   & 0.265±0.005 & 0.312±0.005 & 0.691±0.004 & 26.2±0.2 & 161.0±4.0 & 0.120±0.005 & 4.80±0.14  & 320.3±1.6    \\
Capsule        & Depth   &  Multiple heard   & 0.246±0.003 & 0.302±0.004 & 0.623±0.003 & 24.3±0.4 & 205.5±2.7 & 0.099±0.004 & 5.81±0.27  & 880.2±4.9    \\
ViT-V          & Depth   &  Multiple heard   & 0.012±0.002 & 0.188±0.003 & 0.027±0.003 & 12.6±0.2 & 221.0±6.5 & 0.006±0.001 & 11.58±0.18 & 2180.8±11.2   \\
\hline
%-----------------------------------------------------------------------
Capsule        & Depth   &  Multiple unheard  & 0.178±0.001 & 0.255±0.001 & 0.445±0.001 & 19.1±0.1 & 134.0±3.8 & 0.076±0.001 & 8.72±0.15  & 980.8±6.8  \\
ViTScratch-V   & Depth   &  Multiple unheard  & 0.167±0.001 & 0.232±0.002 & 0.422±0.001 & 18.0±0.1 & 270.5±2.9 & 0.069±0.001 & 9.05±0.12  & 1160.5±3.2   \\
ViT-V          & Depth   &  Multiple unheard  & 0.013±0.001 & 0.170±0.002 & 0.020±0.001 & 10.3±0.1 & 296.6±4.0 & 0.005±0.002 & 14.17±0.25 & 1420.5±6.8   \\
\hline
ViT-V-EM      & Depth  & Telephone  & 0.400±0.003 & 0.456±0.004 & 0.806±0.001 & 30.1±0.1 & 153.9±1.0 & 0.203±0.003 & 2.66±0.10  & 40.1±0.1       \\
Capsule-V       & Depth  & Telephone  & 0.380±0.005 & 0.429±0.003 & 0.803±0.001 & 29.7±0.1 & 156.1±1.0 & 0.185±0.005 & 2.91±0.10  & 100.1±0.5    \\
ViT             & Depth  & Telephone  & 0.370±0.002 & 0.419±0.003 & 0.797±0.001 & 29.3±0.1 & 167.4±1.1 & 0.171±0.003 & 3.05±0.14  & 180.1±0.9    \\
Caps-V-EMul     & Depth  & Telephone  & 0.333±0.003 & 0.391±0.005 & 0.758±0.004 & 28.5±0.4 & 190.2±1.3 & 0.135±0.003 & 3.32±0.13  & 860.1±4.4    \\
ViT-V-EMul      & Depth  & Telephone  & 0.217±0.002 & 0.264±0.001 & 0.643±0.005 & 24.1±0.6 & 221.9±1.0 & 0.099±0.005 & 6.07±0.34  & 340.3±1.9    \\
ViTScratch      & Depth  & Telephone  & 0.019±0.001 & 0.094±0.001 & 0.065±0.005 & 11.4±0.5 & 225.8±1.1 & 0.008±0.001 & 12.30±0.46 & 1681.0±8.9    \\
\hline
\end{tabular}
}
\end{table*}
%----------------------------------------------------------------------------------
%
%----------------------------------------------------------------------------------

\subsection{Visualization of the learned weights from visual encoder} \label{appx: fl-visual}
\begin{figure*}[h!]
	\centering
	\includegraphics[width=0.6\textwidth,height=200pt]{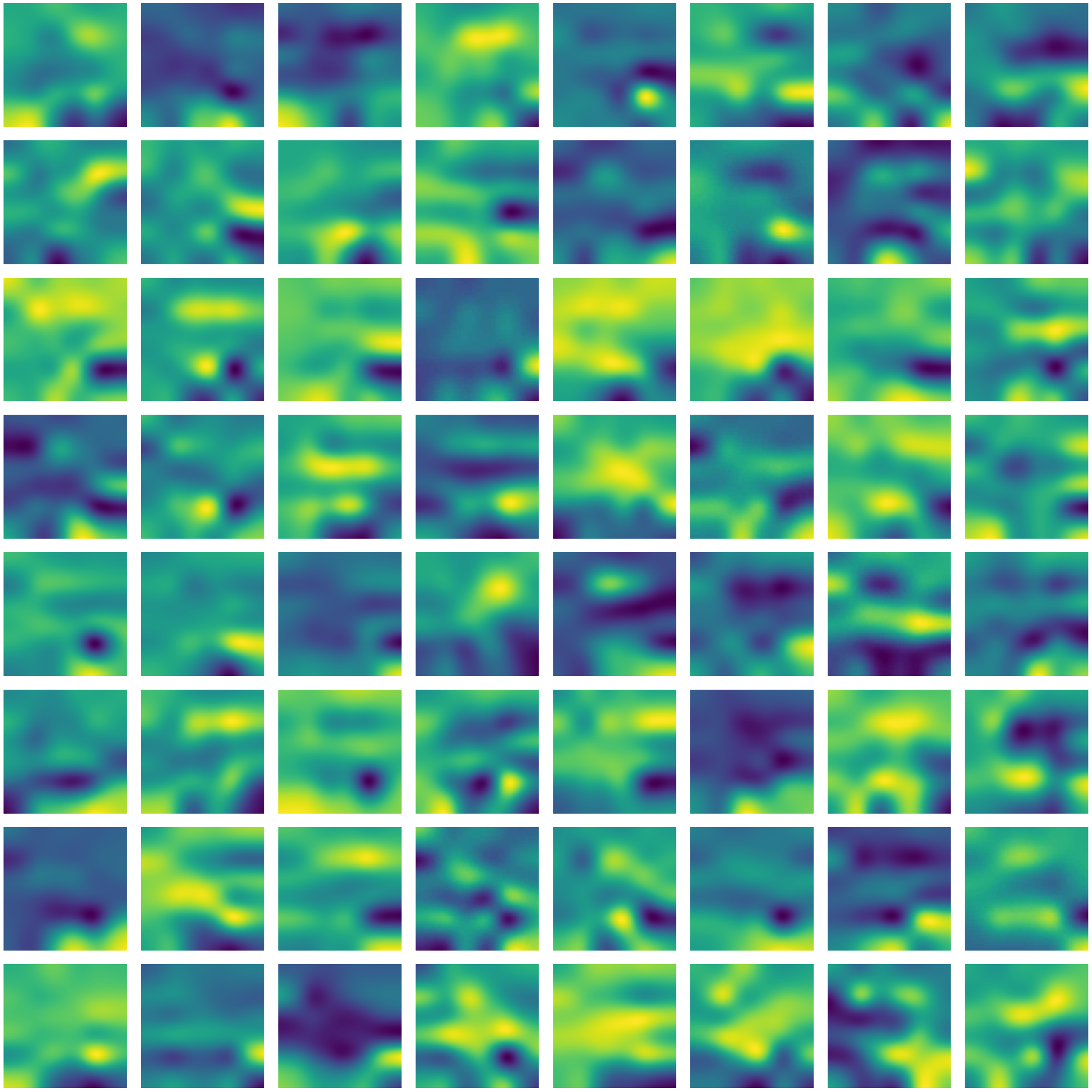}
	\caption{
		Weight visualization for layer 6 (64-channel) of FSAAVN visual encoder.
	}
	\label{fig: fl-visual-l3-64}
\end{figure*}

\begin{figure*}[h!]
\centering
\includegraphics[width=0.6\textwidth,height=200pt]{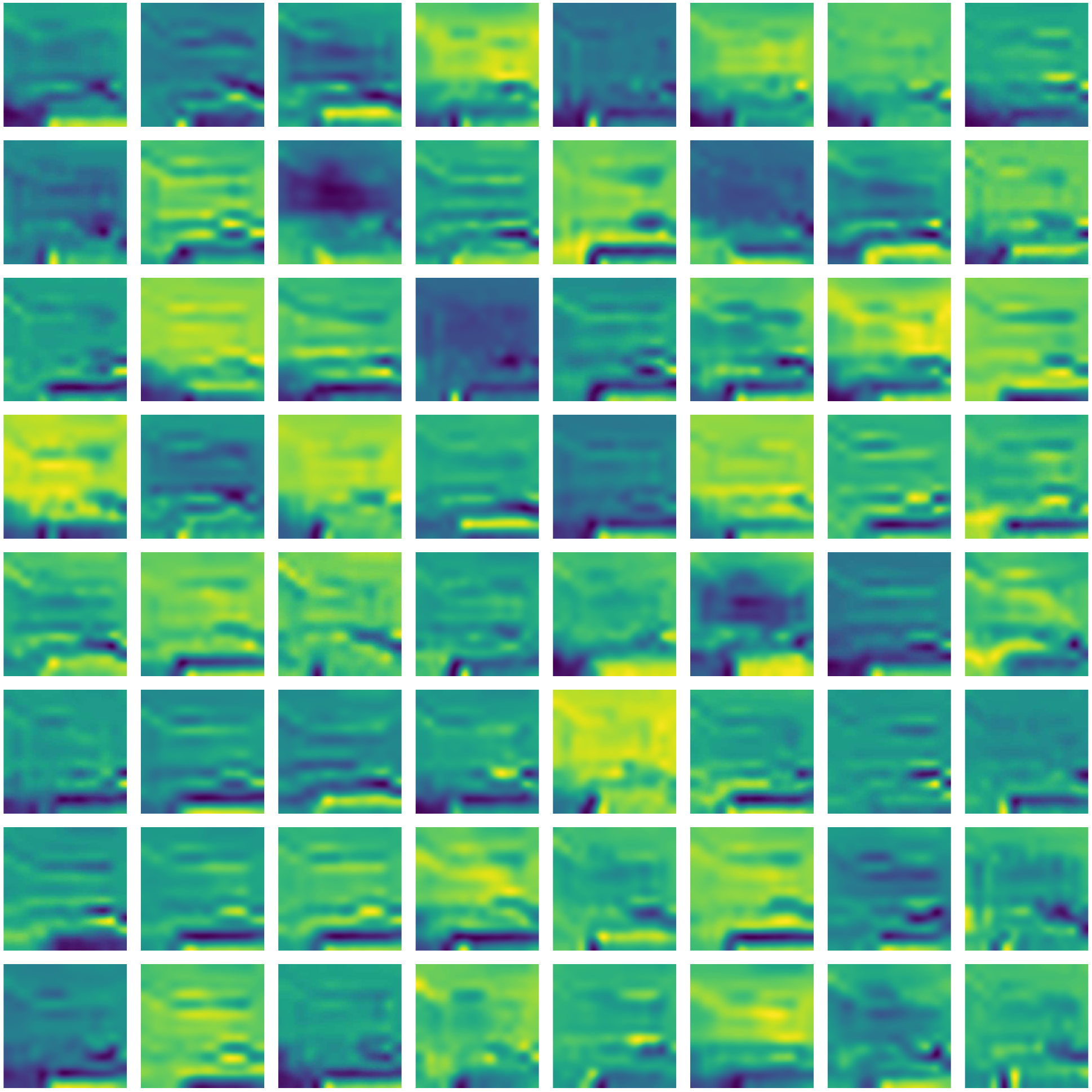}
\caption{
	Weight visualization for layer 4 (32-channel) of FSAAVN visual encoder.
}
\label{fig: fl-visual-l2-32}
\end{figure*}

\begin{figure*}[h!]
\centering
\includegraphics[width=0.6\textwidth,height=100pt]{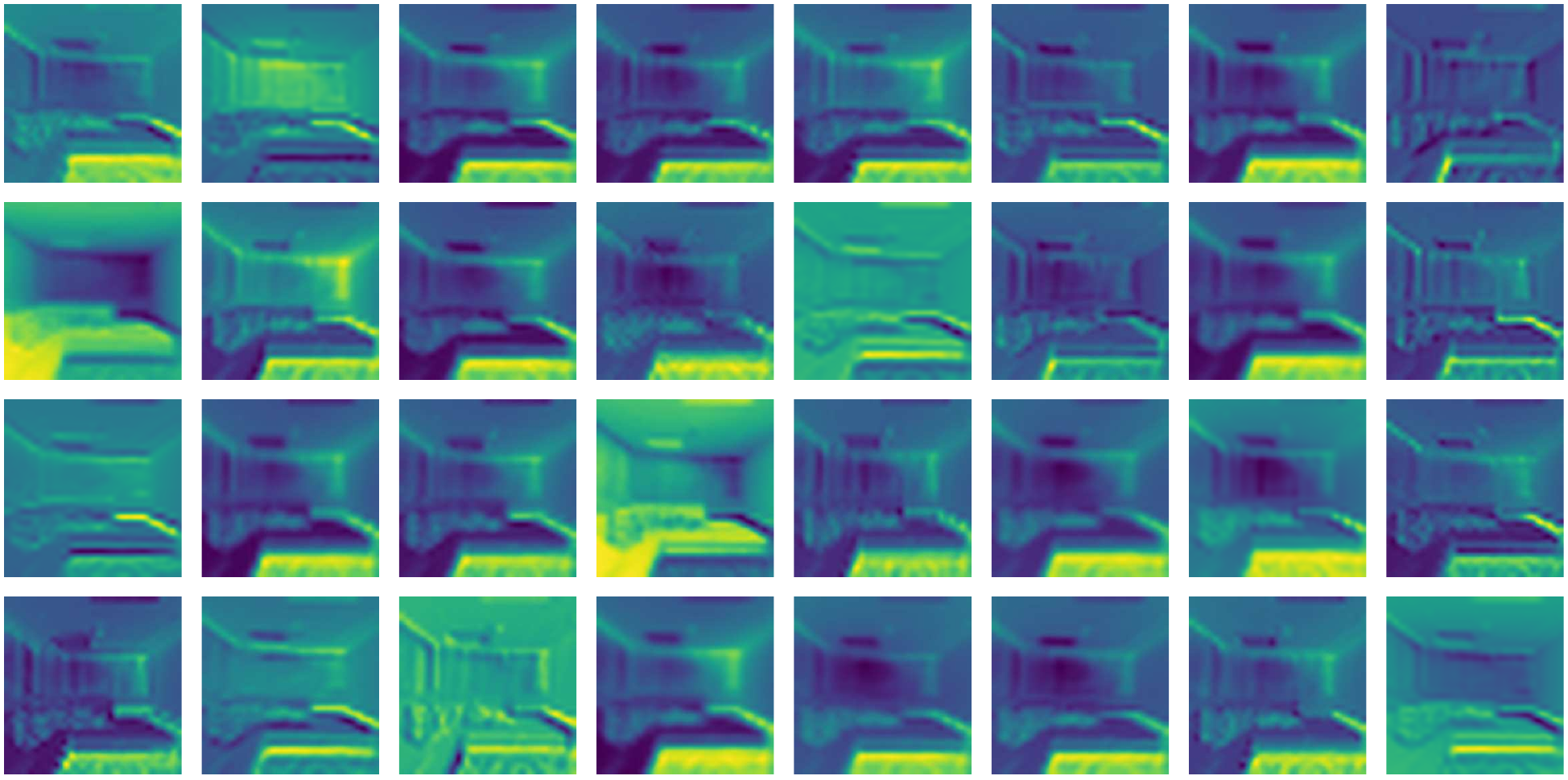}
\caption{
	Weight visualization for layer 2 (32-channel) of FSAAVN visual encoder.
}
\label{fig: fl-visual-l1-32}
\end{figure*}

In this section, we present a visualization of the weight data from each channel and each layer of the FSANN visual encoder.
Fig.\ref{fig: fl-visual-l3-64}, \ref{fig: fl-visual-l2-32} and \ref{fig: fl-visual-l1-32} are weights visualizations for the 6-th, 4-th and 2-nd layer of the FSAAVN visual encoder with 64, 32, 32 channels respectively.

%-----------------------------------------

%-----------------------------------------
%-----------------------------------------
%----------------------------------------------------------------------------------
%
%----------------------------------------------------------------------------------
\clearpage
\subsection{Navigation trajectories using RGBD input} \label{appx: traj-RGBD}
\begin{figure*}[h!]
	\centering
	\includegraphics[width=\textwidth]{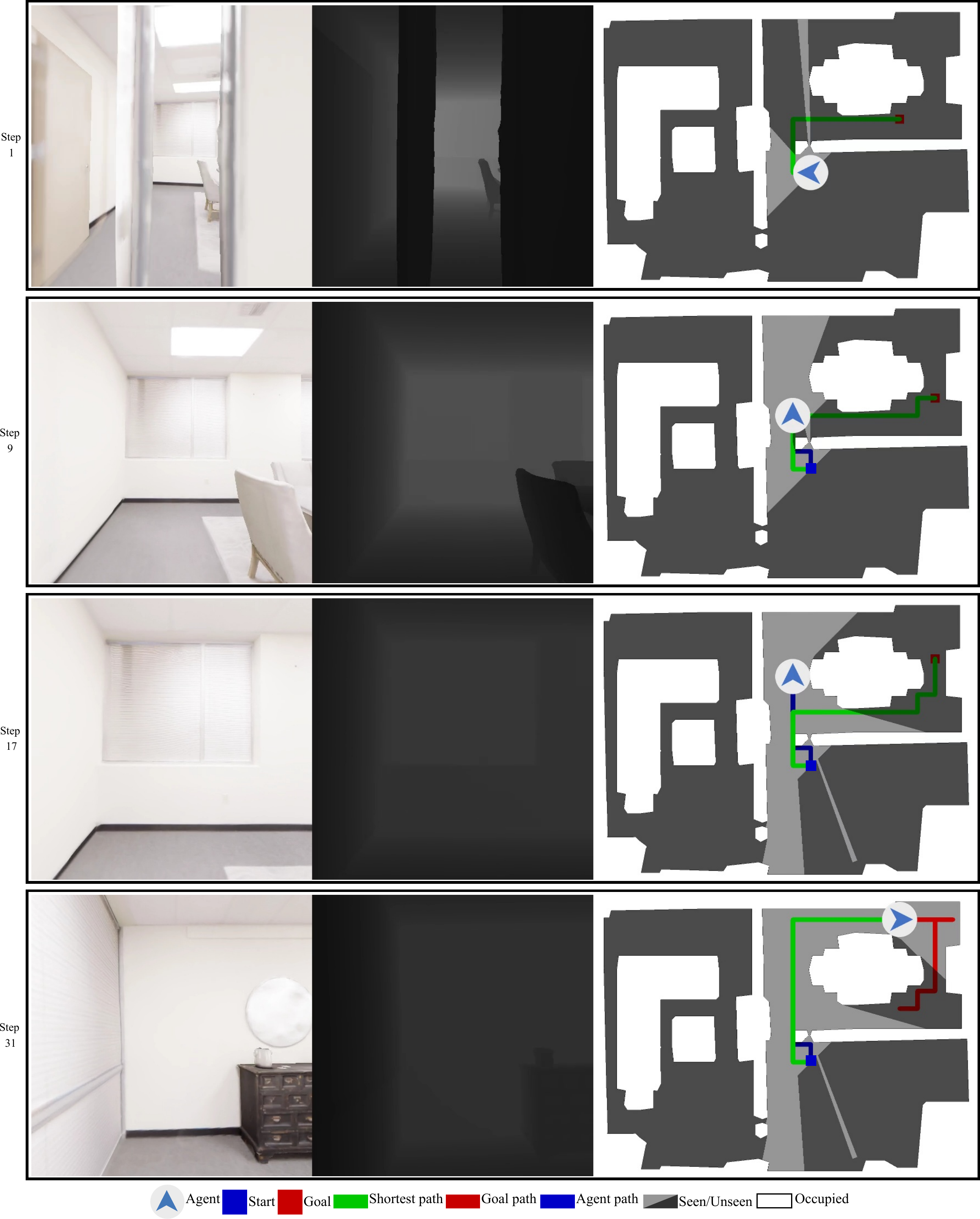}
	\caption{
		A demonstration plot of the trajectories from FSAAVN at steps 1, 9, 17, and 31 (rows). 
		The visual input includes RGB and depth images. This episode is played out on Replica with SPLT=0.88.
	}
	\label{fig: traj-replica-RGBD-FSAAVN-splt0.88-V1}
\end{figure*}

\begin{figure*}[h!]
	\centering
 \vspace{30pt}
	\includegraphics[width=\textwidth]{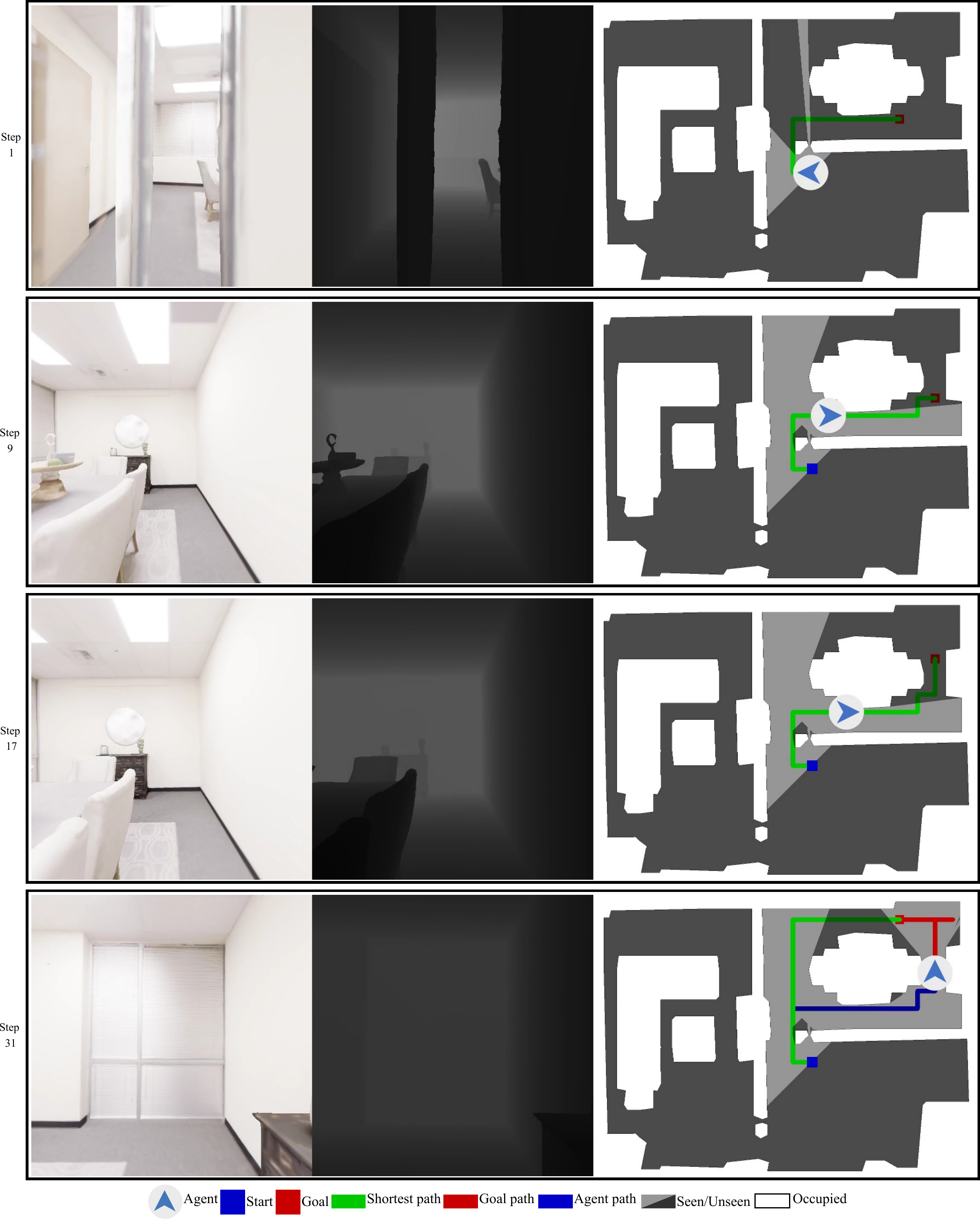}
	\caption{
  A demonstration plot of the trajectories from SoundSpaces at steps 1, 9, 17, and 31 (rows). 
		The visual input includes RGB and depth images. This episode is played out on Replica with SPLT=0.70.
	}
	\label{fig: traj-replica-RGBD-AVN-splt0.70-V1}
\end{figure*}

\begin{figure*}[ht!]
\centering
\includegraphics[width=\textwidth]{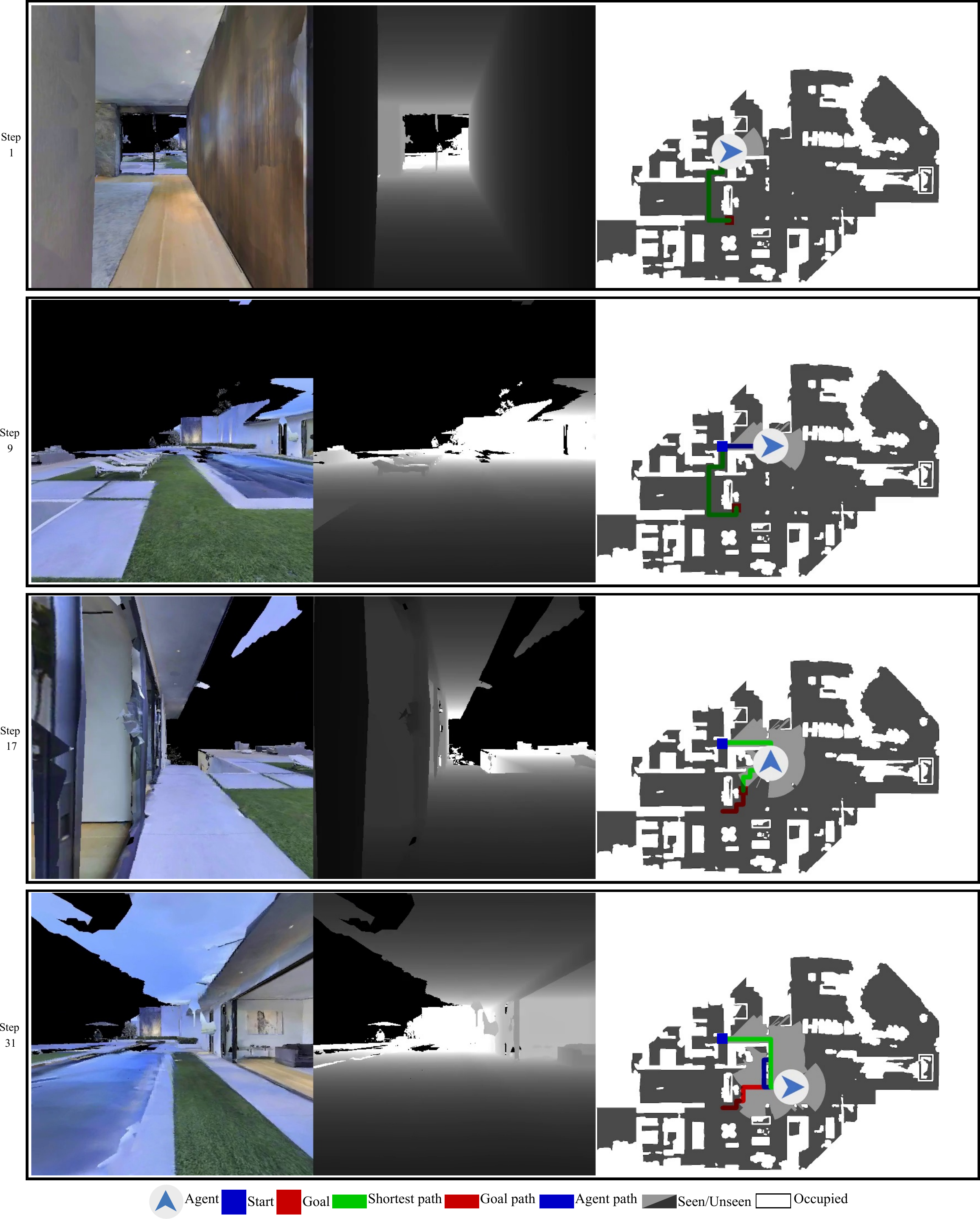}
\caption{
A demonstration plot of the trajectories from FSAAVN at steps 1, 9, 17, and 31 (rows). 
The visual input includes RGB and depth images. This episode is played out on Matterport3D with SPLT=0.89.
}
\label{fig: traj-mp3d-RGBD-FSAAVN-splt0.89-V1}
\end{figure*}

\begin{figure*}[ht!]
\centering
\includegraphics[width=\textwidth]{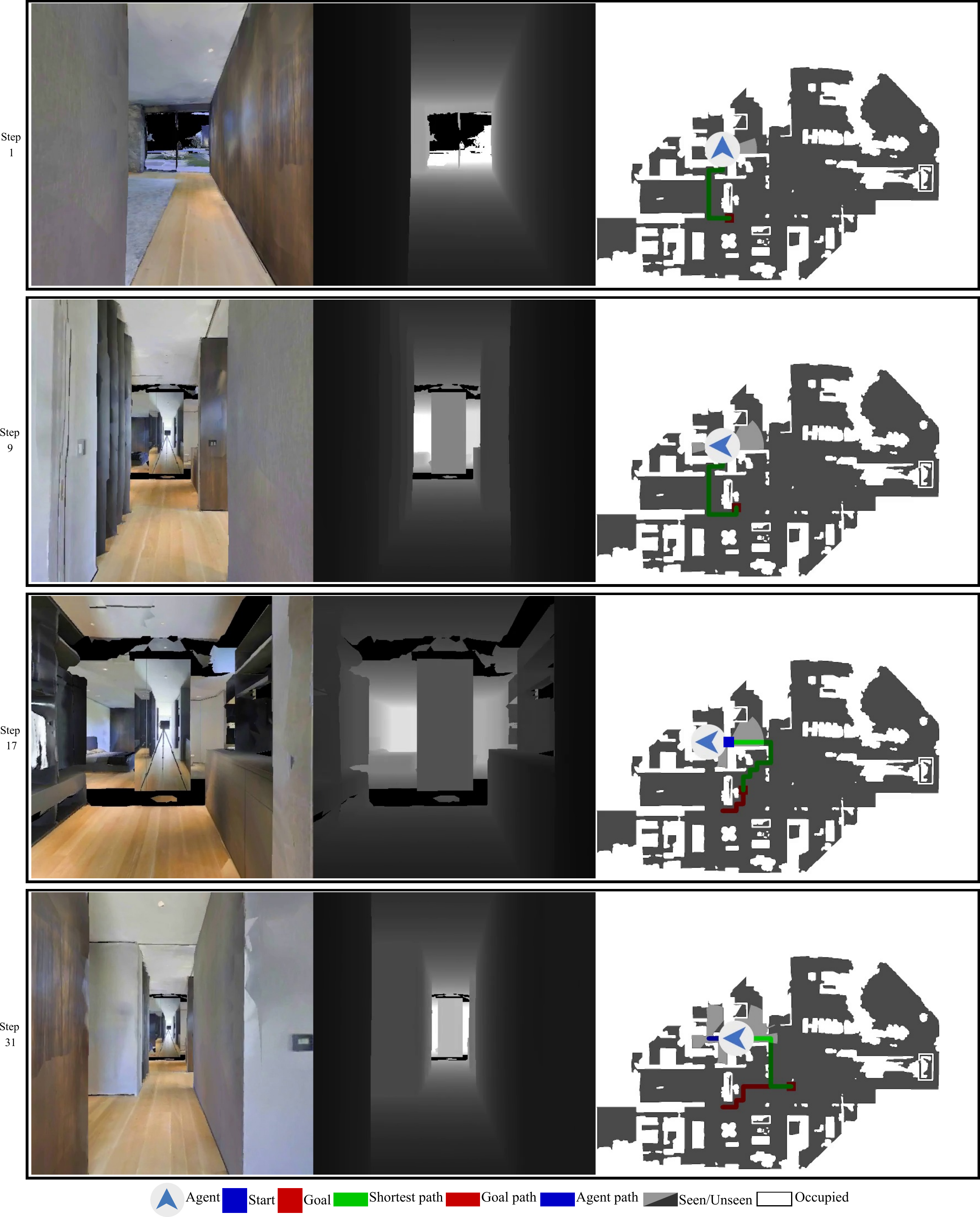}
\caption{\small
A demonstration plot of the trajectories from SoundSpaces at steps 1, 9, 17, and 31 (rows). 
The visual input includes RGB and depth images. This episode is played out on Matterport3D with SPLT=0.67.
}
\label{fig: traj-mp3d-RGBD-AVN-splt0.67-V1}
\end{figure*}

The exemplary plots of the trajectories from \textbf{FSAAVN} and \textbf{SoundSpaces} using both RGB and depth images are shown in Fig.\ref{fig: traj-replica-RGBD-FSAAVN-splt0.88-V1} to \ref{fig: traj-mp3d-RGBD-AVN-splt0.67-V1}.
The rows in each figure represent the sampled steps (1, 9, 17, and 31).
%----------------------------------------------------------------------------------
%
%----------------------------------------------------------------------------------

%----------------------------------------------------------------------------------
%
%----------------------------------------------------------------------------------

%----------------------------------------------------------------------------------
%
%----------------------------------------------------------------------------------

%----------------------------------------------------------------------------------
%
%----------------------------------------------------------------------------------
\clearpage
\subsection{Navigation trajectories using depth input} \label{appx: traj-depth}
\begin{figure*}[ht!]
	\centering
	\includegraphics[width=0.7\textwidth]{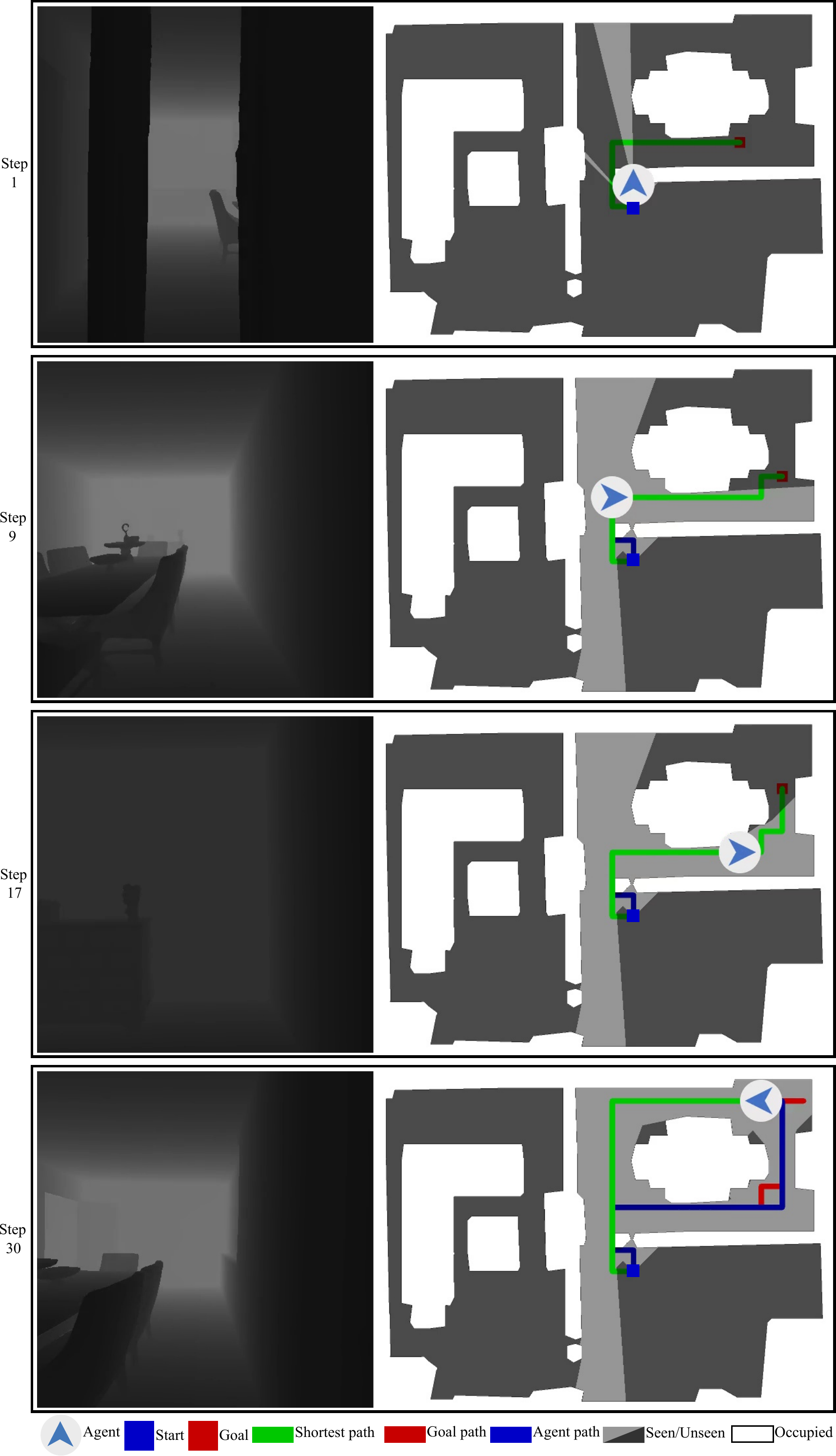}
	\caption{
  A demonstration plot of the trajectories from FSAAVN at steps 1, 9, 17, and 30 (rows). 
The visual input is depth images. This episode is played out on Replica with SPLT=0.89.
	}
	\label{fig: traj-replica-Depth-FSAAVN-splt0.89-V1}
\end{figure*}

\begin{figure*}[ht!]
\centering
\includegraphics[width=0.7\textwidth]{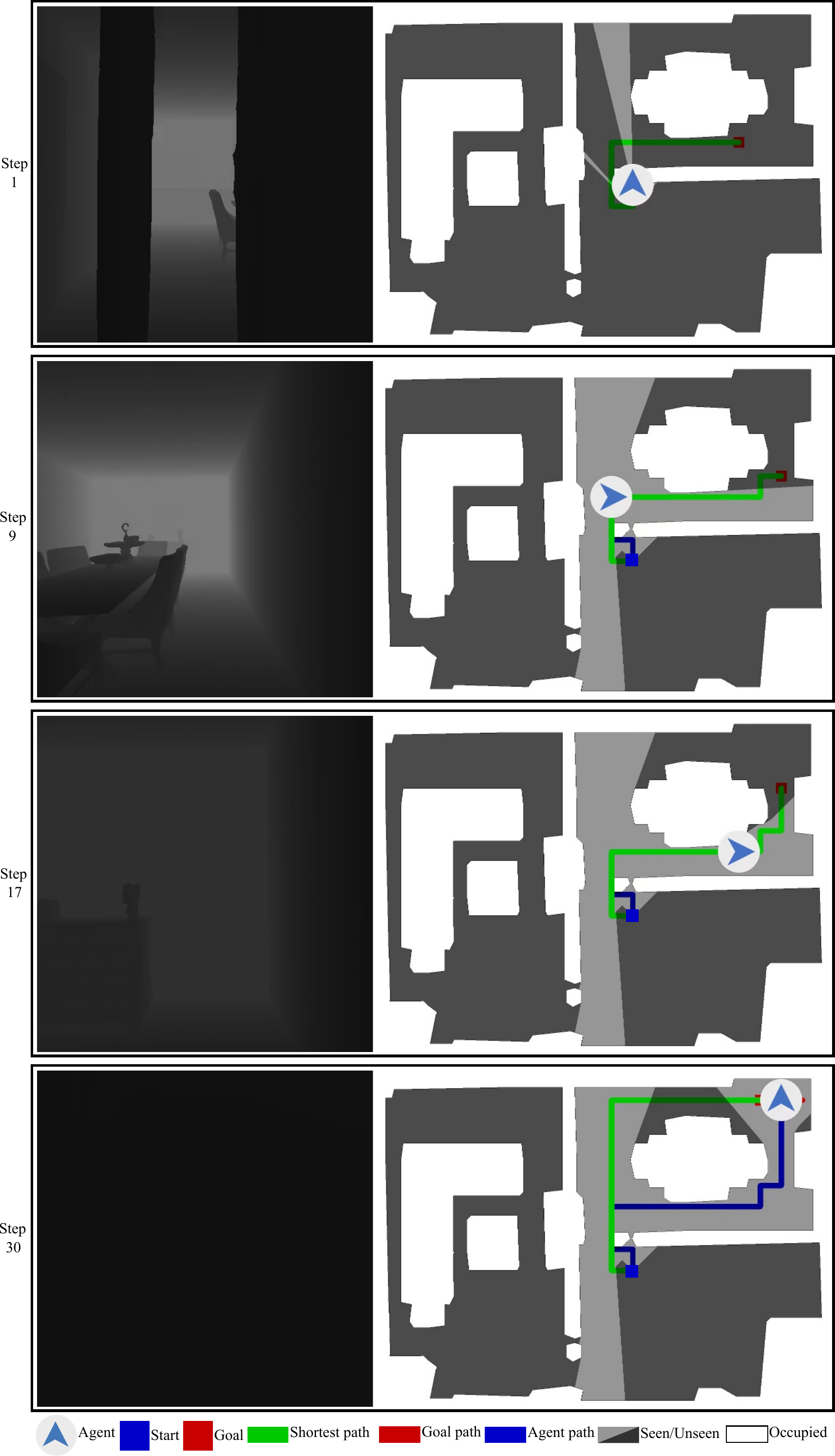}
\caption{
   A demonstration plot of the trajectories from SoundSpaces at steps 1, 9, 17, and 30 (rows). 
The visual input is depth images. This episode is played out on Replica with SPLT=0.79.
}
\label{fig: traj-replica-Depth-AVN-splt0.79-V1}
\end{figure*}

\begin{figure*}[ht!]
\centering
\includegraphics[width=0.7\textwidth]{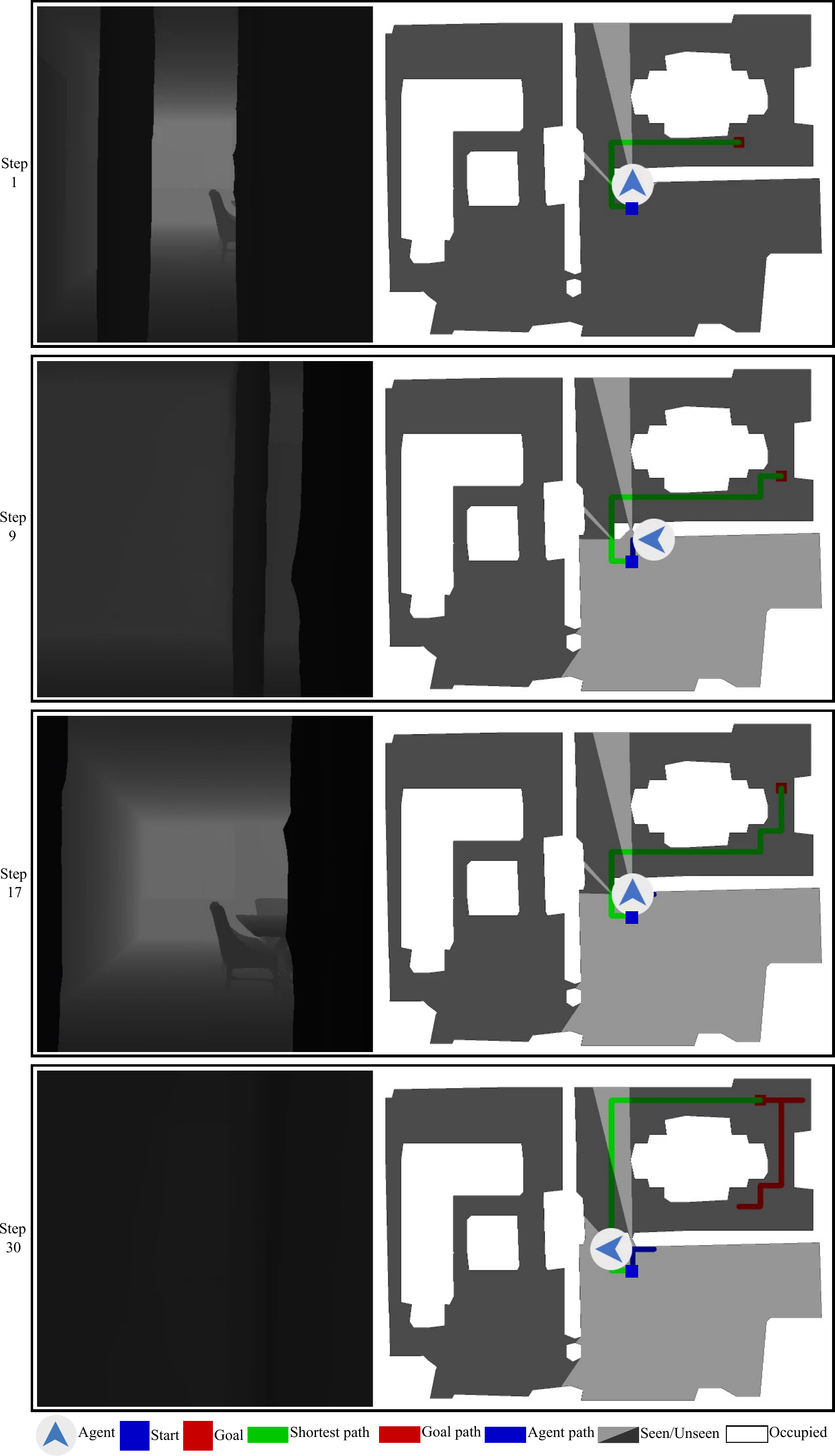}
\caption{
    A demonstration plot of the trajectories from SoundSpaces-EMul at steps 1, 9, 17, and 30 (rows). 
The visual input is depth images. This episode is played out on Replica with SPLT=0.68.
}
\label{fig: traj-replica-Depth-DotMul-splt0.68-V1}
\end{figure*}

\begin{figure*}[ht!]
\centering
\includegraphics[width=0.7\textwidth]{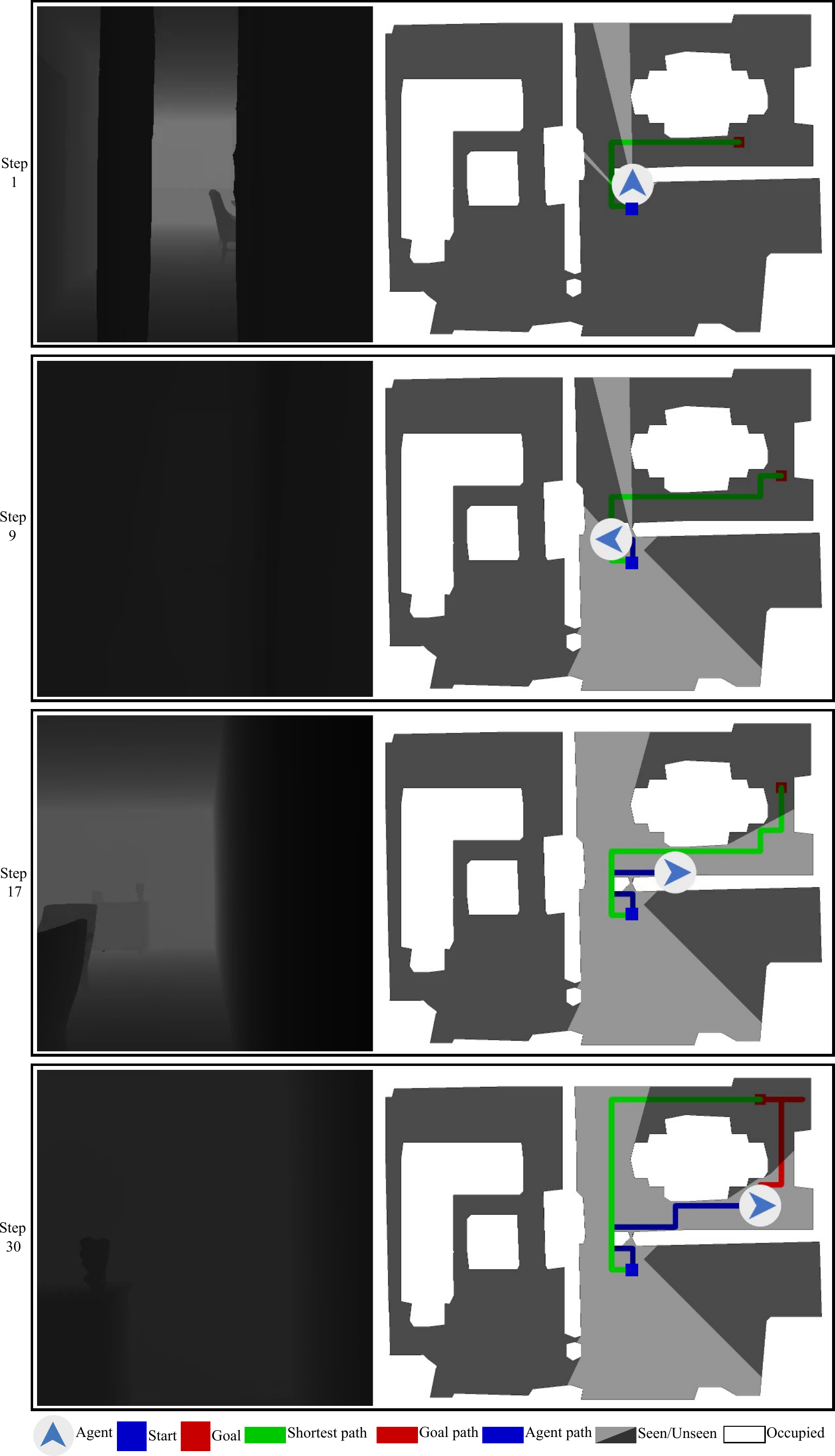}
\caption{
A demonstration plot of the trajectories from SoundSpaces-EM at steps 1, 9, 17, and 30 (rows). 
The visual input is depth images. This episode is played out on Replica with SPLT=0.62.
}
\label{fig: traj-replica-Depth-Mean-splt0.62-V1}
\end{figure*}

\begin{figure*}[ht!]
\centering
\includegraphics[width=0.7\textwidth]{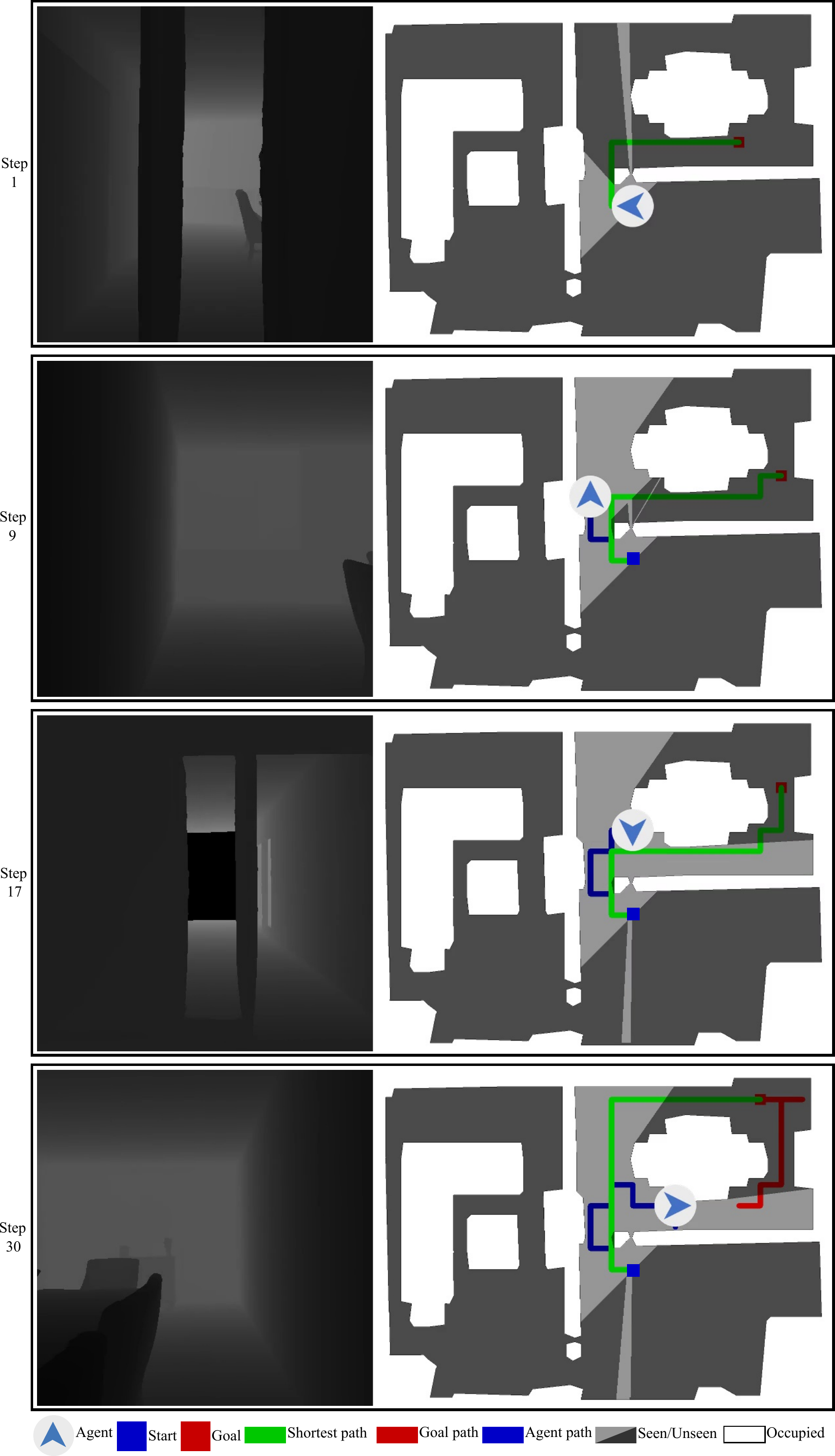}
\caption{
A demonstration plot of the trajectories from CMHM at steps 1, 9, 17, and 30 (rows). 
The visual input is depth images. This episode is played out on Replica with SPLT=0.29.
}
\label{fig: traj-replica-Depth-CMHM-splt0.29-V1}
\end{figure*}

\begin{figure*}[ht!]
\centering
\includegraphics[width=0.7\textwidth]{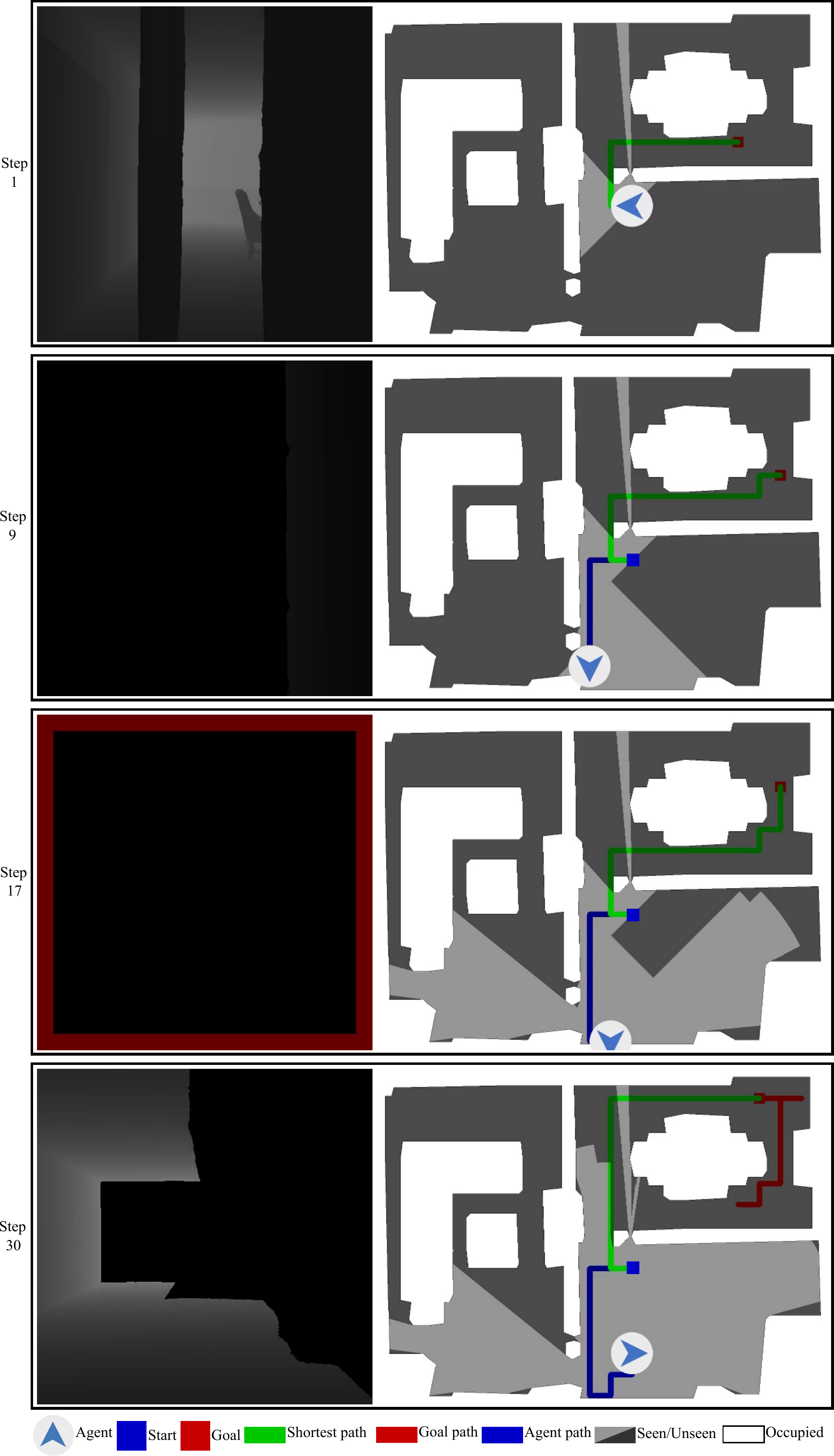}
\caption{
A demonstration plot of the trajectories from AV-WaN at steps 1, 9, 17, and 30 (rows). 
The visual input is depth images. This episode is played out on Replica with SPLT=0.17.
}
\label{fig: traj-replica-Depth-WAN-splt0.17-V1}
\end{figure*}

\begin{figure*}[ht!]
\centering
\includegraphics[width=0.7\textwidth]{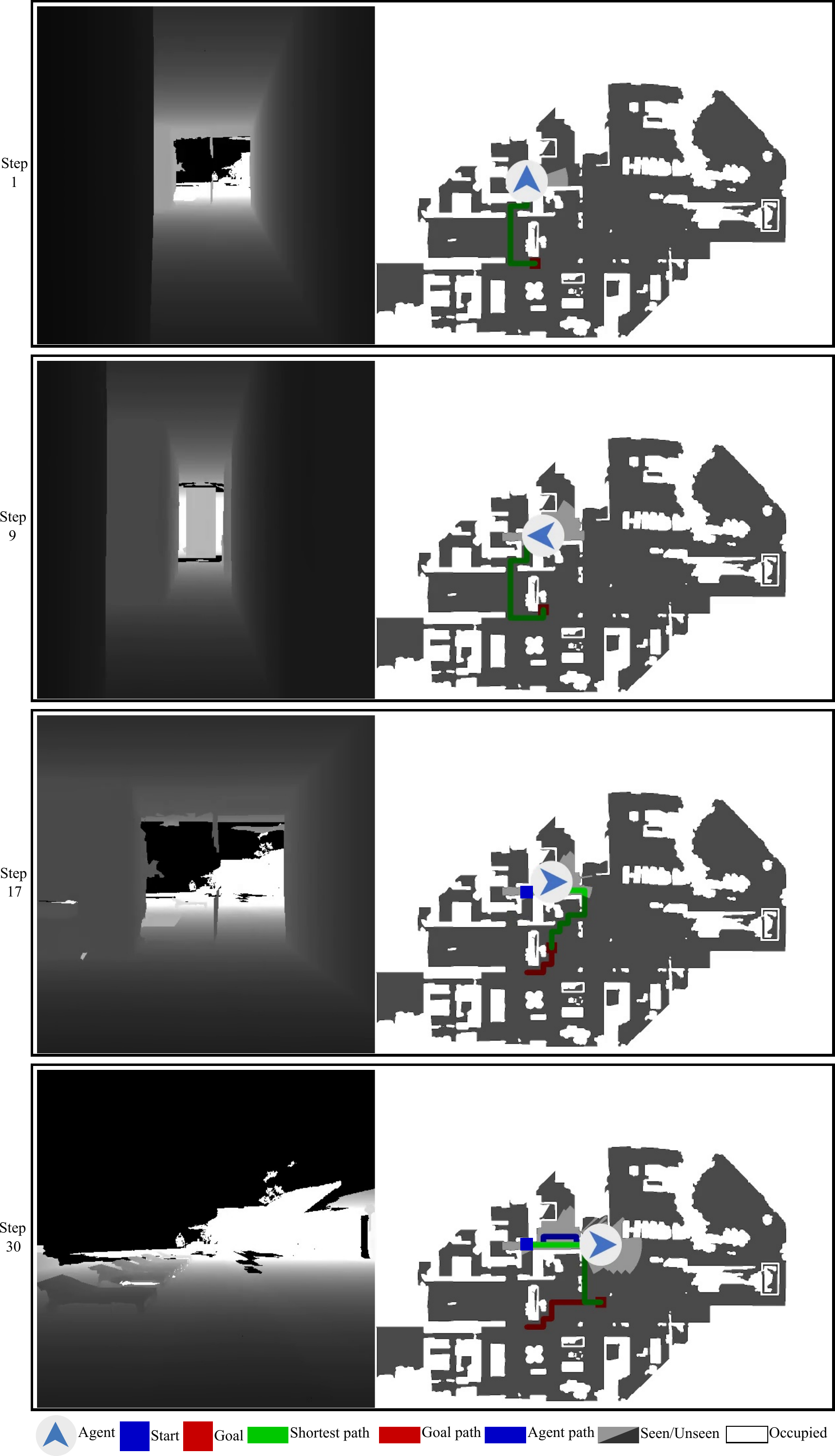}
\caption{
A demonstration plot of the trajectories from FSAAVN at steps 1, 9, 17, and 30 (rows). 
The visual input is depth images. This episode is played out on Matterport3D with SPLT=0.90.
}
\label{fig: traj-mp3d-Depth-FSAAVN-splt0.90-V1}
\end{figure*}

\begin{figure*}[ht!]
\centering
\includegraphics[width=0.7\textwidth]{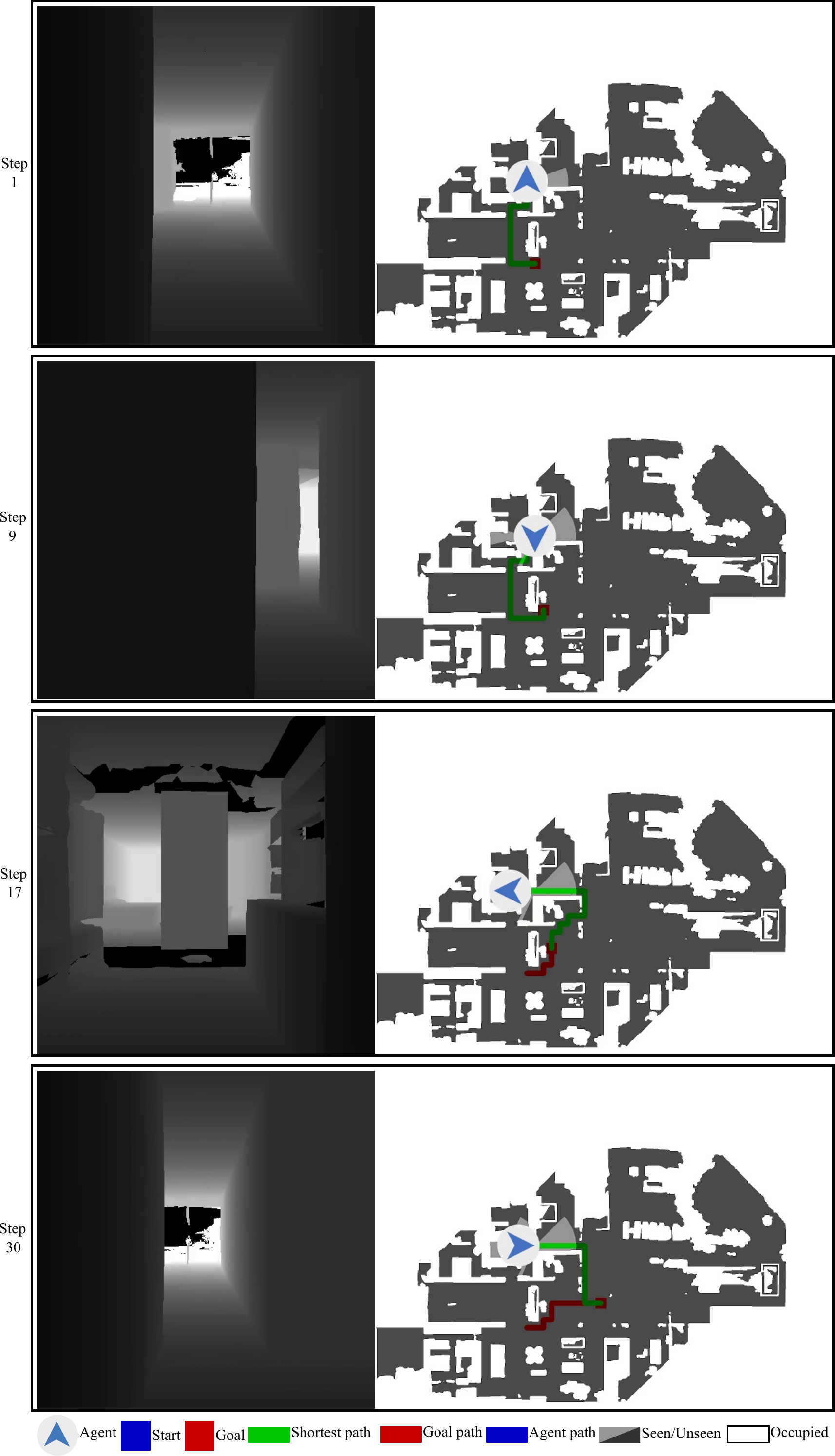}
\caption{
A demonstration plot of the trajectories from SoundSpaces at steps 1, 9, 17, and 30 (rows). 
The visual input is depth images. This episode is played out on Matterport3D with SPLT=0.66.
}
\label{fig: traj-mp3d-Depth-AVN-splt0.66-V1}
\end{figure*}

\begin{figure*}[ht!]
\centering
\includegraphics[width=0.7\textwidth]{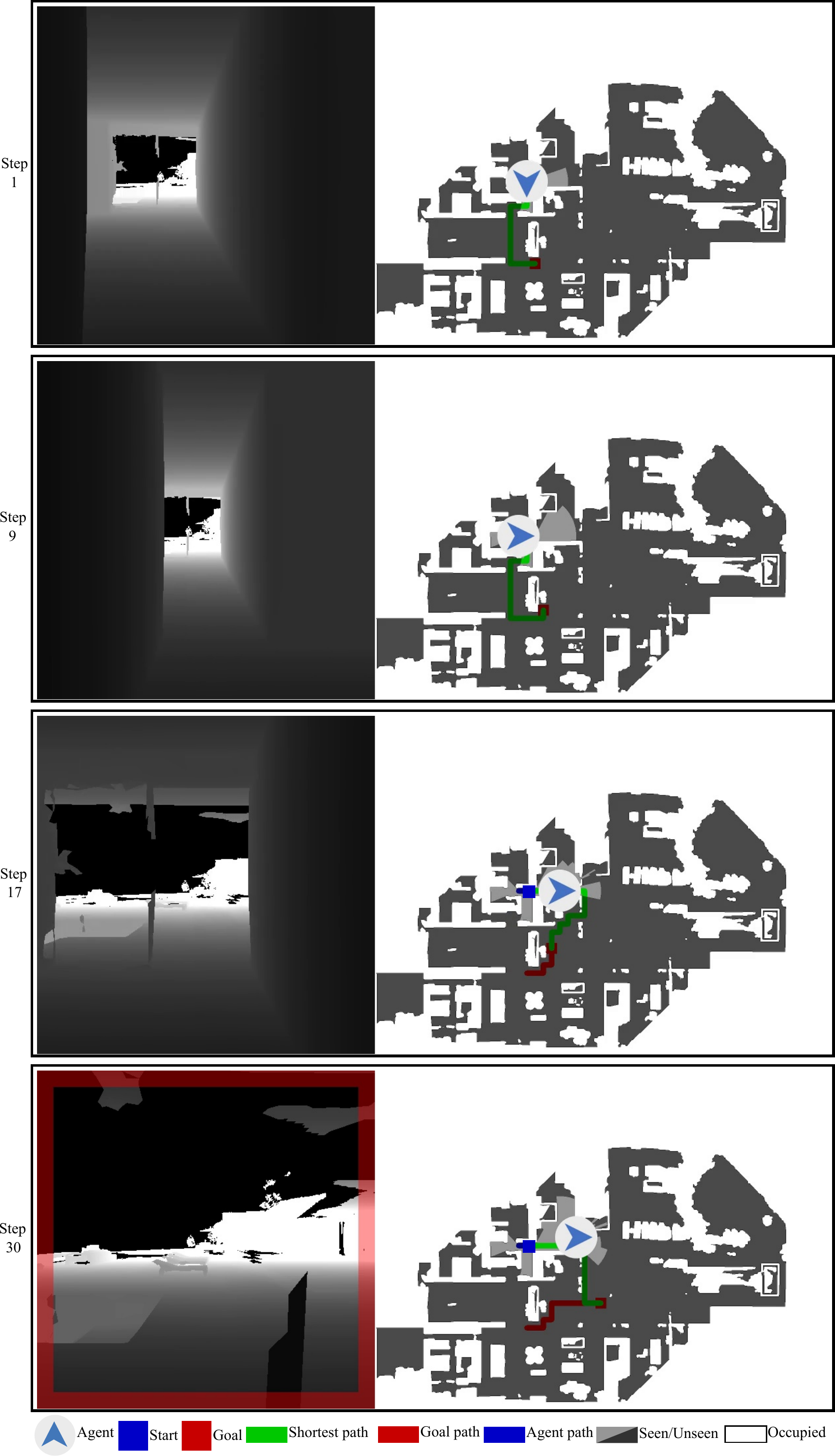}
\caption{
A demonstration plot of the trajectories from SoundSpaces-EMul at steps 1, 9, 17, and 30 (rows). 
The visual input is depth images. This episode is played out on Matterport3D with SPLT=0.76.
}
\label{fig: traj-mp3d-Depth-Dotmul-splt0.76-V1}
\end{figure*}

\begin{figure*}[ht!]
\centering
\includegraphics[width=0.7\textwidth]{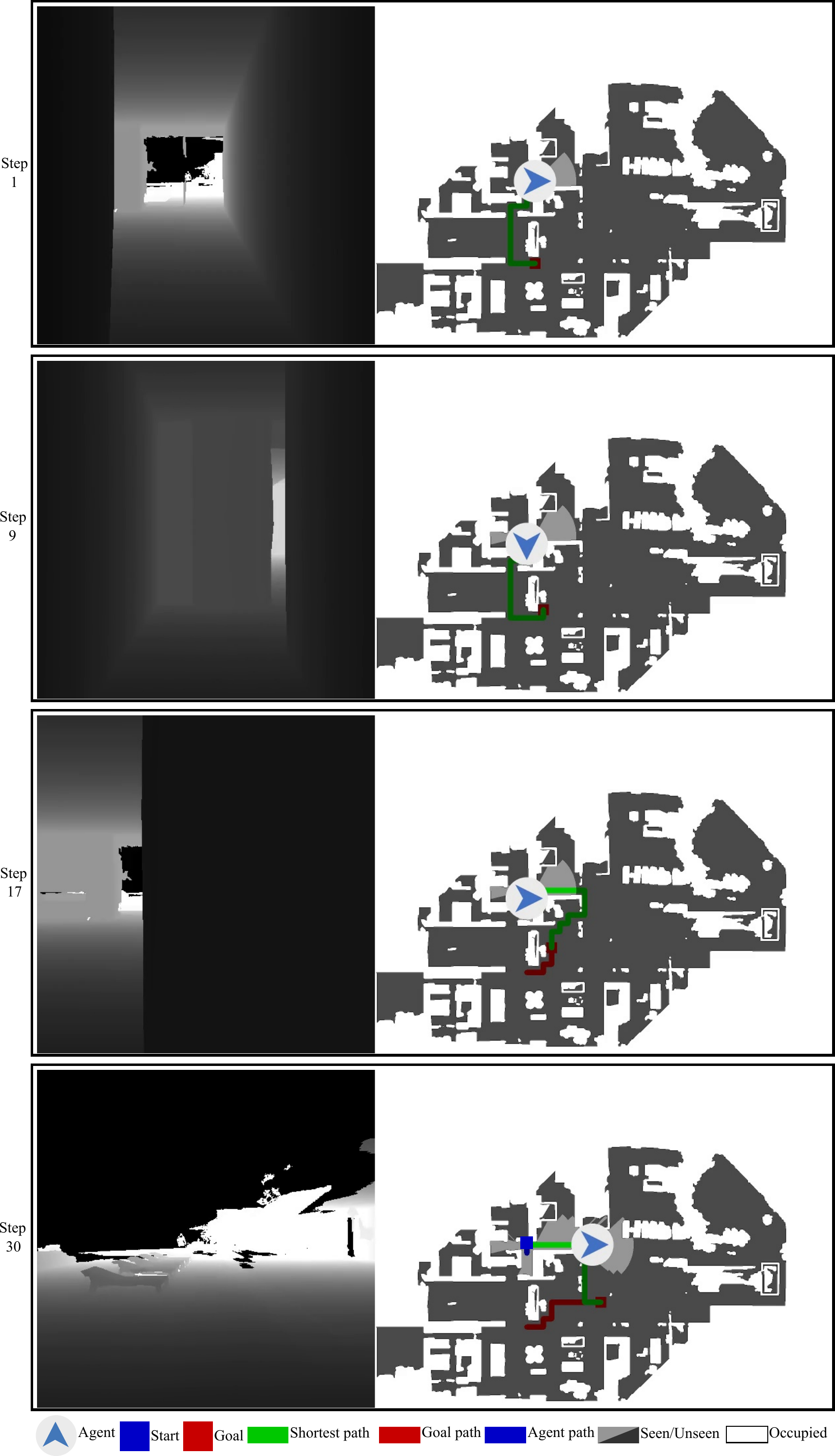}
\caption{
A demonstration plot of the trajectories from SoundSpaces-EM at steps 1, 9, 17, and 30 (rows). 
The visual input is depth images. This episode is played out on Matterport3D with SPLT=0.82.
}
\label{fig: traj-mp3d-Depth-Mean-splt0.82-V1}
\end{figure*}

\begin{figure*}[ht!]
\centering
\includegraphics[width=0.7\textwidth]{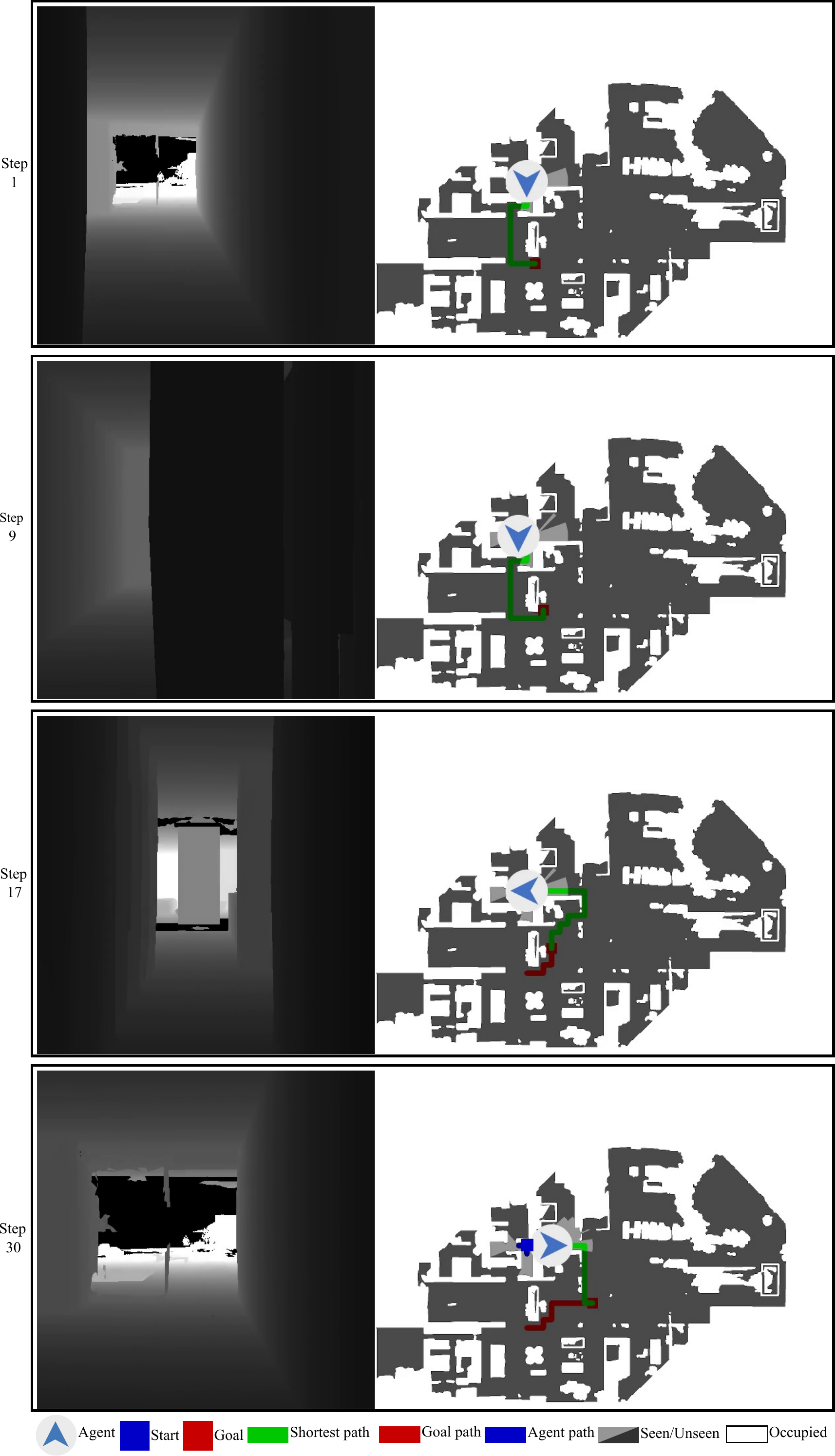}
\caption{
A demonstration plot of the trajectories from CMHM at steps 1, 9, 17, and 30 (rows). 
The visual input is depth images. This episode is played out on Matterport3D with SPLT=0.56.
}
\label{fig: traj-mp3d-Depth-CMHM-splt0.56-V1}
\end{figure*}

\begin{figure*}[ht!]
\centering
\includegraphics[width=0.7\textwidth]{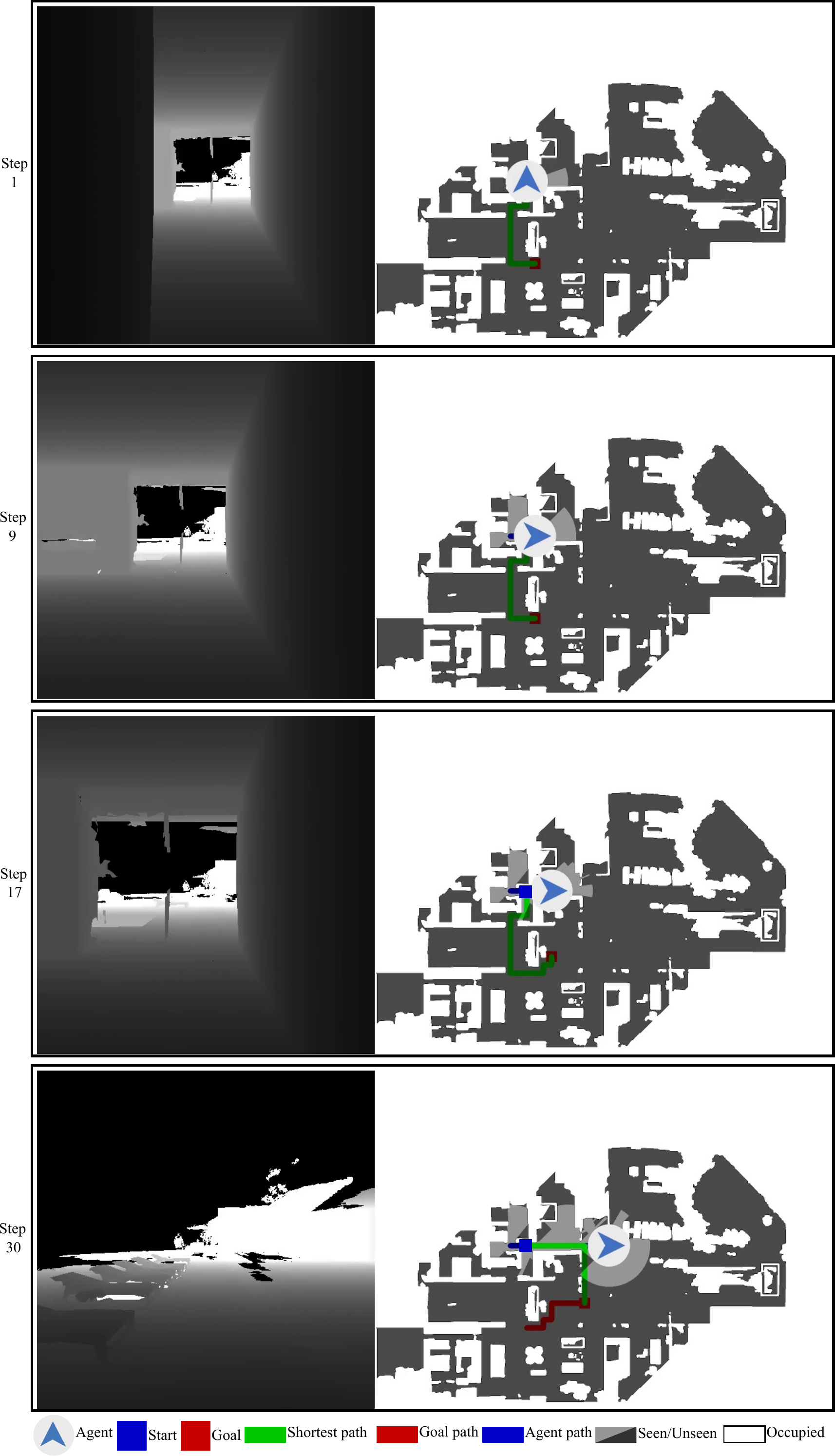}
\caption{
A demonstration plot of the trajectories from AV-WaN at steps 1, 9, 17, and 30 (rows). 
The visual input is depth images. This episode is played out on Matterport3D with SPLT=0.34.
}
\label{fig: traj-mp3d-Depth-WAN-splt0.34-V1}
\end{figure*}

The plots of the trajectories from different methods (FSAAVN, SoundSpaces, SoundSpaces-EMul, SoundSpaces-EM, CMHM and AV-WaN) using only depth images are shown in Fig.\ref{fig: traj-replica-Depth-FSAAVN-splt0.89-V1} to \ref{fig: traj-mp3d-Depth-WAN-splt0.34-V1}.
The rows in each figure represent the sampled steps (1, 9, 17, and 30).

%----------------------------------------------------------------------------------
%
%----------------------------------------------------------------------------------

%----------------------------------------------------------------------------------
%
%----------------------------------------------------------------------------------

%----------------------------------------------------------------------------------
%
%----------------------------------------------------------------------------------

%----------------------------------------------------------------------------------
%
%----------------------------------------------------------------------------------

%----------------------------------------------------------------------------------
%
%----------------------------------------------------------------------------------

%----------------------------------------------------------------------------------
%
%----------------------------------------------------------------------------------

%----------------------------------------------------------------------------------
%
%----------------------------------------------------------------------------------

%----------------------------------------------------------------------------------
%
%----------------------------------------------------------------------------------

%----------------------------------------------------------------------------------
%
%----------------------------------------------------------------------------------

%----------------------------------------------------------------------------------
%
%----------------------------------------------------------------------------------

%----------------------------------------------------------------------------------
%
%----------------------------------------------------------------------------------

%----------------------------------------------------------------------------------
%
%----------------------------------------------------------------------------------

\clearpage
\section{Example navigation video clips}\label{appendix: videos}
We also provide example navigation video clips, which can be downloaded from the bottom of the web page\footnote{https://yyf17.github.io/FSAAVN/index.html} in the section called ``Example navigation video clips''.
% \url{https://yyf17.github.io/FSAAVN/index.html}(navigation to Example navigation video clips in the web)
% The use of the video is described in the file README.md in the root directory of the supplementary material.
The file names of these video clips are listed in Table~\ref{appx: demos}.
Each video file is named following the format of ``\textbf{xx}-\textbf{yy}-\textbf{zz}-splt0.\textbf{mm}.mp4'', 
where \textbf{xx} denotes the dataset, $\textbf{xx} \in \{ \text{Replica}, \text{Matterport3D}\}$;
\textbf{yy} represents vision type, $\textbf{yy} \in \{ \text{Depth}, \text{RGBD}\}$;
\textbf{zz} indicate the method, $\textbf{zz} \in \{$ SoundSpaces, SoundSpaces-EMul, SoundSpaces-EM, FSAAVN, CMHM, AV-WaN $\}$;
and \textbf{mm} is the value of SPLT metric.
%-------------------------------
\begin{table*}[h]
\centering
\caption{Specification of example navigataion video clips.}
\label{appx: demos}
\vspace{0.1in}
% \resizebox{0.9\linewidth}{!}{
\begin{tabular}{ll}
\hline
Id & File name  \\
\hline
1	&	\small	Matterport3D-Depth-AV-WaN-splt0.34.mp4	\\
2	&	\small	Matterport3D-Depth-CMHM-splt0.56.mp4	\\
3	&	\small	Matterport3D-Depth-FSAAVN-splt0.90.mp4	\\
4	&	\small	Matterport3D-Depth-SoundSpaces-EM-splt0.82.mp4	\\
5	&	\small	Matterport3D-Depth-SoundSpaces-EMul-splt0.76.mp4	\\
6	&	\small	Matterport3D-Depth-SoundSpaces-splt0.66.mp4	\\
7	&	\small	Matterport3D-RGBD-FSAAVN-splt0.89.mp4	\\
8	&	\small	Matterport3D-RGBD-SoundSpaces-splt0.67.mp4	\\
9	&	\small	Replica-Depth-AV-WaN-splt0.17.mp4	\\
10	&	\small	Replica-Depth-CMHM-splt0.29.mp4	\\
11	&	\small	Replica-Depth-FSAAVN-splt0.89.mp4	\\
12	&	\small	Replica-Depth-SoundSpaces-EM-splt0.62.mp4	\\
13	&	\small	Replica-Depth-SoundSpaces-EMul-splt0.68.mp4	\\
14	&	\small	Replica-Depth-SoundSpaces-splt0.79.mp4	\\
15	&	\small	Replica-RGBD-FSAAVN-splt0.88.mp4	\\
16	&	\small	Replica-RGBD-SoundSpaces-splt0.70.mp4	\\
\hline
\end{tabular}
% }
\end{table*}
%-------------------------------------------
% \section{Ethics statement. } \label{appendix: ethics}
% This research does NOT involve any human subject. Our dataset is not related to any issue of privacy and can be used publicly. All authors of this paper follow the \href{https://bmvc2022.org/authors/submit-your-paper/}{BMVC's ethics guidelines}. 
%----------------------------------------------------------------------------------
%
%----------------------------------------------------------------------------------
%\clearpage

\EnableMainstart 
\else
    % \bibliography{egbib}
    \end{document}